\documentclass[fleqn,usenatbib]{mnras}

\usepackage{newtxtext,newtxmath}

\usepackage[T1]{fontenc}
\usepackage{ae,aecompl}

\usepackage{graphicx}	
\usepackage{amsmath}	
\usepackage{amssymb}	
\usepackage{pdflscape}	
\usepackage{enumitem}   

\newcommand{\met}{\mbox{12+$\log$(O/H)}}

\newcommand{\hii}{\mbox{H\,{\scshape ii}}}
\newcommand{\Ha}{\mbox{H$\alpha$}}
\newcommand{\Hb}{\mbox{H$\beta$}}
\newcommand{\Hg}{\mbox{H$\gamma$}}
\newcommand{\Hd}{\mbox{H$\delta$}}
\newcommand{\oiia}{\mbox{[O\,{\scshape ii]}}\,$\lambda$3727}

\newcommand{\oii}{\mbox{[O\,{\scshape ii]}}\,$\lambda$3727,29}
\newcommand{\oiiia}{\mbox{[O\,{\scshape iii]}}\,$\lambda$4959}
\newcommand{\oiiib}{\mbox{[O\,{\scshape iii]}}\,$\lambda$5007}
\newcommand{\nii}{\mbox{[N\,{\scshape ii]}}\,$\lambda$6585}
\newcommand{\neiii}{\mbox{[Ne\,{\scshape iii}]}\,$\lambda$3869}

\title[Resolved Metallicity at $z\approx$1]{Resolved scaling relations and metallicity gradients on sub-kiloparsec scales at $z\approx1$}

\author[Patr\'icio et al.]{V. Patr\'icio$^{1,2}$,\thanks{E-mail: vera.patricio@dark-cosmology.dk}
J. Richard$^{2}$,
D. Carton$^{2}$,
C. P\'eroux$^{3,4}$,
T. Contini$^{5}$,
J. Brinchmann$^{6,7}$,
\newauthor
J. Schaye$^{7}$,
P. M. Weilbacher$^{8}$,
T. Nanayakkara$^{7}$,
M. Maseda$^{7}$,
G. Mahler$^{9}$,
and
\newauthor
L. Wisotzki$^{8}$\\
$^{1}$ DARK, Niels Bohr Institute, University of Copenhagen, Lyngbyvej 2, 2100 Copenhagen, Denmark\\
$^{2}$ Univ Lyon, Univ Lyon1, Ens de Lyon, CNRS, Centre de Recherche Astrophysique de Lyon UMR5574, F-69230, Saint-Genis-Laval, France\\
$^{3}$ European Southern Observatory (ESO), Karl-Schwarzschild-Str.2, D-85748 Garching b. M\"unchen, Germany \\
$^{4}$ Aix Marseille Univ, CNRS, LAM, Laboratoire d'Astrophysique de Marseille, Marseille, France \\
$^{5}$ Institut de Recherche en Astrophysique et Plan\'etologie (IRAP), Universit\'e de Toulouse, CNRS, UPS, F-31400 Toulouse, France\\
$^{6}$ Instituto de Astrof\'isica e Ci\^encias do Espa\c{c}o, Universidade do Porto, CAUP, Rua das Estrelas, 4150-762 Porto,
Portugal\\
$^{7}$ Leiden Observatory, Leiden University, PO Box 9513, 2300 RA Leiden, The Netherlands\\
$^{8}$ Leibniz-Institut f\"ur Astrophysik Potsdam (AIP), An der Sternwarte 16, 14482 Potsdam, Germany\\
$^{9}$ Department of Astronomy, University of Michigan, 1085 South University Ave, Ann Arbor, MI 48109, USA}

\date{Accepted XXX. Received YYY; in original form ZZZ}

\pubyear{2019}

\begin{document}
\label{firstpage}
\pagerange{\pageref{firstpage}--\pageref{lastpage}}
\maketitle

\begin{abstract}
The existence of a spatially resolved Star-Forming Main Sequence (rSFMS) and a spatially resolved Mass-Metallicity  Relation (rMZR) is now well established for local galaxies. Moreover, gradients with metallicity decreasing with radius seem to be common in local disc galaxies. These observations suggest that galaxy formation is a self-regulating process, and provide constraints for galaxy evolution models. Studying the evolution of these relations at higher redshifts is still however very challenging.
In this paper, we analyse three gravitationally lensed galaxies at $z=$ 0.6, 0.7 and 1, observed with MUSE and SINFONI. These galaxies are highly magnified by galaxy clusters, which allow us to observe resolved scaling relations and metallicity gradients on physical scales of a couple of hundred parsecs, comparable to studies of local galaxies. We confirm that the rSFMS is already in place at these redshifts on sub-kpc scales, and establish, for the first time, the existence of the rMZR at higher redshifts. 
We develop a forward-modelling approach to fit 2D metallicity gradients of multiply imaged lensed galaxies in the image plane, and derive gradients of -0.027$\pm$0.003, -0.019$\pm$0.003 and -0.039$\pm$0.060 dex/kpc. Despite the fact that these are clumpy galaxies, typical of high redshift discs, the metallicity variations in the galaxies are well described by global linear gradients, and we do not see any difference in metallicity associated with the star-forming clumps.
\end{abstract}





\section{Introduction}
\label{sec:intro}

It has now been well established that the masses, star-formation rates, and gas metallicities of star-forming galaxies are tightly correlated by two relations: the \textit{Star-Forming Main Sequence} (SFMS), that relates stellar mass and star-formation rates, and the \textit{Mass-Metallicity Relation} (MZR), relating mass and metallicity. These scaling relations have been observed from $z=0$ up to $z=6$ \citep[e.g.][]{Brinchmann2004,Tremonti2004,Erb2006,Whitaker2012,Speagle2014}. It has even been argued that these three properties are connected by a single plane, the \textit{fundamental mass-metallicity relation} \citep{Lara2010,Mannucci2010}, that does not evolve with redshift, although its existence is still controversial  \citep[e.g.][]{Sanchez2013,Erroz-Ferrer2019}. 

These scaling relations are well explained by "reservoir" models. In these analytical models, after an initial phase of gas accretion, galaxies self-regulate their star-formation rates, evolving in a quasi-steady state \citep[e.g.][]{Schaye2010,Dutton2010,Bouche2010,Dave2012,Lilly2013}. These models can successfully predict the SFMS and MZR using only a couple of fairly simple "regulators" of star-formation, such as gas infall rates, outflow rates, and gas recycling rates, without involving any details about star-forming processes. Since it is possible to reproduce these scaling relations without specifying any details on star-forming or stellar feedback processes, additional observables are needed to further our understanding of galaxy evolution.

In recent years, with the increasing number of Integral Field Unit (IFU) spectrograph surveys of local disc galaxies (e.g. CALIFA \citealt{Sanchez2014}, MaNGA \citealt{Bundy2015}, SAMI \citealt{Croom2012} and MAD \citealt{Erroz-Ferrer2019}), it has also been established that both the Star-Forming Main Sequence and the Mass-Metallicity Relation exist on sub-galactic scales \citep[e.g.][]{Sanchez2013,Erroz-Ferrer2019}. Moreover, it has been argued that these resolved relations are in fact more fundamental than the integrated ones \citep{Rosales-Ortega2012,Barrera-Ballesteros2016}, i.e., that the galaxy wide scaling-relations are a consequence of the local relations between stellar mass surface density, star-formation surface density and metallicity. 

It is unclear if the reservoir models can be extended to explain these resolved relations, since they are based on isolated galactic systems, rather then contiguous and possibly interacting kpc-scale regions (but see, for example, \citealt{Ho2015,Carton2015,DEugenio2018}). It is also not clear what the reservoir would correspond to in this case and how the equilibrium phase would be reached. New and additional observables are needed to advance these simple but powerful models of galaxy evolution, as well as to test complex simulations that include sub-grid recipes for smaller-scale physical processes \citep[e.g.][]{Trayford2018}.

Another area of rapid development thanks to recent IFU surveys is the study of metallicity gradients. In the local Universe, disc galaxies are commonly observed to have higher metallicities in the centre than in the outskirts (a negative metallicity gradient) \citep[e.g.][]{Pilyugin2015,Ho2015,Carton2015,Sanchez-Menguiano2016,Belfiore2017}, possibly with a universal slope when normalised to the galaxy size \citep[e.g.][]{Sanchez2014,Ho2015}. The negative metallicity gradients can be explained by the 'inside-out' disc growth scenario, where the inner parts of galaxies are formed at earlier times and are, consequently, more metal enriched than the outskirts \citep{Larson1976}. However, models that predict metallicity gradients compatible with the ones observed locally, make different predictions for gradients at earlier epochs, predicting either a steepening of the gradient at earlier epochs \citep[e.g.][]{Pilkington2012}, or a flattening \cite[e.g.][]{Mott2013}. 

Deriving metallicity gradients at high-$z$ remains challenging. While in the local Universe metallicity gradients are generally negative, at high-$z$ a wide range of gradients, from negative to positive, has been measured. \cite{Wuyts2016} measured the metallicity gradients of star-forming galaxies at $z=0.6-2.7$, finding that they are, on average, flat. At slightly lower redshifts, $z=0.1-0.8$, \cite{Carton2018} find a negative median gradient, but with a large scatter (8\% of their sample have significant positive gradients and 31\% are consistent with flat gradients).

The evolution of the resolved scaling relations with cosmic time also remains difficult to probe, since it requires both a high signal-to-noise ratio and a high spatial resolution. \cite{Wuyts2013}, and more recently \cite{Abdurrouf2018}, have measured the resolved Star-Forming Main Sequence on kilo-parsec scales in massive galaxies (M$_\star$ > 10$^{10}$ M$_\odot$) at $z=0.7-1.8$ using multi-band high-resolution HST imaging, finding that the rSFMS was already in place at those redshifts. On the other hand, the resolved Mass-Metallicity Relation has still not been studied outside the local Universe until now.

In this work, we combine IFU optical and IR data from MUSE \citep{Bacon2010} and SINFONI \citep{Eisenhauer2003} observations of strongly gravitationally lensed arcs at z$\approx$1 to derive metallicity using multiple line-ratio diagnostics, and dust-corrected SFR from emission lines at physical scales of only a couple of hundreds parsecs. Using these data, we probe the metallicity gradients, resolved Star-Forming Main Sequence and the resolved Mass-Metallicity Relation of typical z$\approx$1 star-forming disc galaxies.

We analyse a sample of 3 strongly lensed galaxies in the background of the Abell S1063/RXJ2248-4431 (AS1063), Abell 370 (A370) and MACSJ1206.2-0847 (M1206) lensing clusters. These gravitational arcs were selected for their large size in the image plane (i.e. as seen in the sky). Despite their high magnification, these galaxies are quite typical of $z=1$ rotating discs. We have presented their basic properties derived from MUSE and HST data in a previous paper, \cite{Patricio2018}. Here, we combine MUSE and SINFONI data to derive the resolved metallicity maps for three of those objects.

This paper is organised as follows. In Section~\ref{sec:data} we present the MUSE and SINFONI data used in this work. In Section~\ref{sec:metallicity} we describe our method to derive metallicity and extinction from line fluxes. In Section~\ref{sec:local_fmz} we analyse the local scaling relations and in Section~\ref{sec:fit} we derive the resolved metallicity maps and describe how we account for lensing. We discuss and summarise our results in Section~\ref{sec:conclusions}. 

Throughout this paper, we adopt a $\Lambda$-CDM cosmology with $\Omega$=0.7, $\Omega_m$=0.3 and H$_0$ = 70 km\,s$^{-1}$\,Mpc$^{-1}$. We adopt a solar metallicity of $12+\log$(O/H) = 8.69 \citep{AllendePrieto2001} and the \cite{Chabrier2003} stellar initial mass function. 

\begin{table*}
\caption{Sample properties derived by \citet{Patricio2018} using MUSE and HST data. From left to right: instrument, observation program identification, point spread function FWHM measured using a Moffat profile, redshift, stellar mass,  magnification-corrected star-formation rate from dust-corrected Balmer lines and effective radius, calculated from the disc length (R$_d$) measured in \citet{Patricio2018} (table 2) using the F160W HST source plane images as R$_e$=1.67835 R$_d$.}
\label{tab:sample}
\centering
\tabcolsep=0.20cm
\begin{tabular}{|lccccccccc|} 
\hline
Object        & $\alpha$ & $\delta$   & Inst. & Program ID & PSF & $z$ & $\log_{10}$ M$_\star$  & SFR$_{\mathrm{MUSE}}$  & R$_e$ \\
         	  &J2000& J2000 & & & ["] &  &[M$_\odot$] & [M$_\odot$/yr] & [kpc]  \\
\hline\hline
AS1063-arc   & 22:48:42  & -44:31:57 & MUSE    & 060.A-9345$^{(a)}$    &  1.03 & 0.6115  & 10.94$\pm$0.05 & 41.5$\pm$4.0 & 7.7$\pm$0.2\\
A370-sys1    & 02:39:53  & -01:35:05 & MUSE    & 094.A-0115, 096.A-0710 &  0.70 & 0.7251  & 10.40$\pm$0.02 & 3.1$\pm$0.3 & 12.0$\pm$0.7\\
M1206-sys1   & 12:06:11  & -08:48:05 & SINFONI & 087.A-0700 	& 0.78 & 1.0366  & 10.90$\pm$0.06 & 107.3$\pm$30.7 & 11.1$\pm$0.2\\
\hline
\end{tabular}
\flushleft $^{(a)}$ see also \citealt{Karman2015} 
\end{table*}

\begin{figure}
	\includegraphics[width=0.5\textwidth]{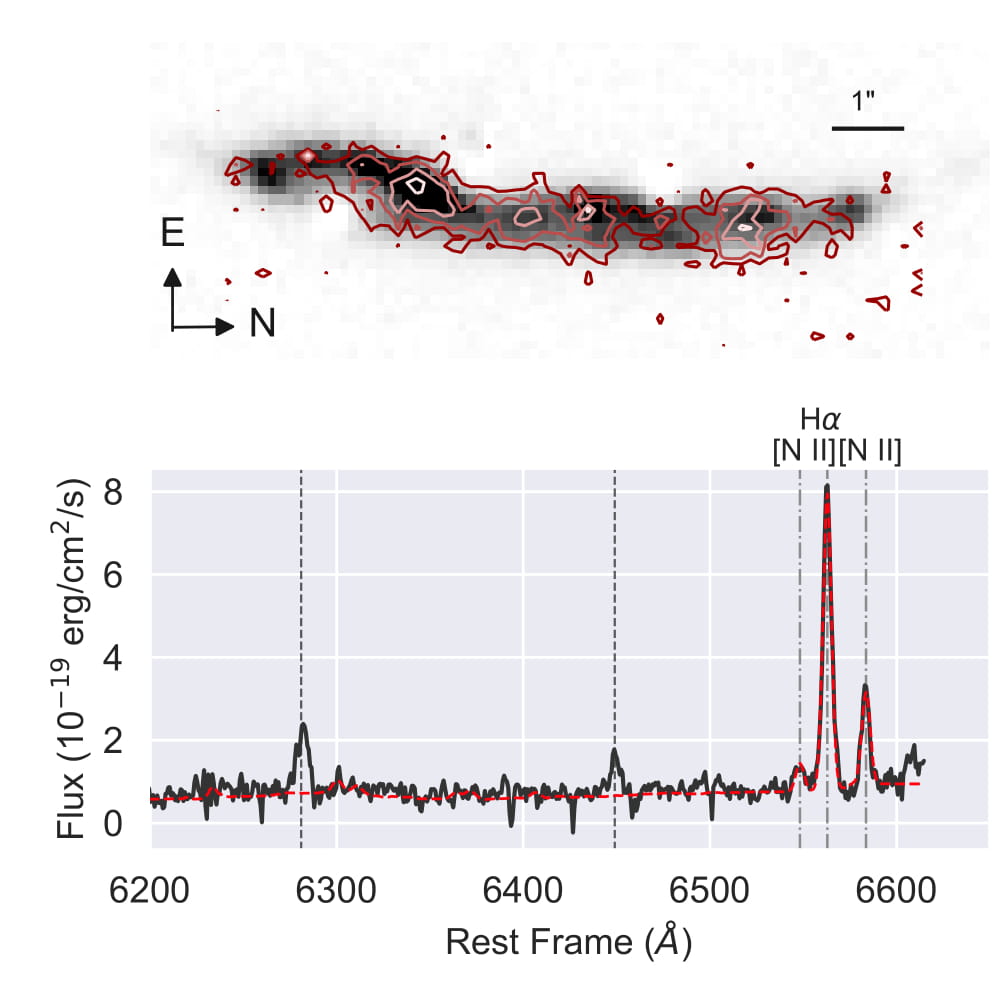}
    \caption{M1206-sys1 data. Top: MUSE \oii\, pseudo-narrow band in grey scale with SINFONI \Ha\, pseudo-narrow band image over-plotted in red contours (surface brightness of 2, 3, 4, 5 $\times$ 10$^{-19}$ erg/cm$^2$/s/arcsec$^2$). Both have been continuum subtracted. Bottom: SINFONI integrated spectrum in black and fit performed with the line fitting code {\sc alfa} \citep{Wesson2016} in dashed red, with the \mbox{[N\,{\sc II]}}\, doublet and \Ha\, identified with dashed-dotted lines and strong sky residuals in dashed lines.}
    \label{fig:macs1206_ha}
\end{figure}

\section{Sample and Data Reduction}
\label{sec:data}

The physical properties of the three galaxies analysed in this work were derived in a homogeneous way from HST and MUSE data in \cite{Patricio2018} (see Table~\ref{tab:sample} for a summary). They have redshifts between 0.6 and 1.0, stellar masses around 10$^{10}$ M$_\odot$, and are compatible with the Fundamental Mass-Metallicity relation \citep{Lara2010,Mannucci2010} up to 0.1 dex. The stellar masses were derived fitting multiple HST bands and the MUSE integrated spectra using {\sc prospector}\footnote{{\sc prospector} (\url{https://doi.org/10.5281/zenodo.1116491})}, a SED fitting code, \cite{Conroy2009} stellar models and the \cite{Chabrier2003} initial mass function. Emission lines were masked during this fit. Dust corrected star-formation rates were calculated from emission lines from the MUSE data, making use of H$\gamma$ or $H\beta$. From the kinematic analysis of the \oii\, emission, we concluded that these are typical rotating discs, representative of the population of star-forming galaxies at $z\approx1$.

For the two lowest redshift galaxies analysed here, AS1063-arc and A370-sys1 (lensed galaxies in the clusters AS1063 and A370), we use MUSE data to derive metallicity maps from optical emission lines (\oii, \oiia, \Hg\,, \Hd\, and \oiiib ). M1206-sys1 was also observed with MUSE, and its global metallicity can be derived from the integrated spectrum using \neiii\, and \oii\, emission lines. However, \neiii\, is too faint to derive the resolved metallicity of this galaxy using MUSE data, and we use instead \Ha\, and \nii\, emission from SINFONI data.

We cannot rule out the presence of an AGN in any of these three galaxies, since none has all the required emission lines to compare it with widely used criteria such as the BPT diagram. However, as we argued in \cite{Patricio2018}, none of these galaxies has {\sc[Mg II]} emission, and the emission lines are generally narrow, particularly at the centre, which makes the presence of broad-line AGNs unlikely, although not impossible.

\subsection{Optical IFU data}
\label{subsec:muse_data}

The MUSE data and their reduction, were already presented in \cite{Patricio2018} and we provide here only a short summary. AS1063 and A370 were observed for 3.25 and 6 hours, respectively. We used the ESO MUSE reduction pipeline version 1.2 \citep{Weilbacher2016} with the usual calibrations (bias, flat, illumination and twilight). The pipeline sky subtraction was improved by using the \textit{Zurich Atmosphere Purge} tool ({\sc ZAP} version 1; \citealt{Soto2016}), a principal component analysis that isolates and removes sky line residuals, on the individual data cubes. 

To determine the Point Spread Function (PSF), the final cubes were compared with \emph{HST} data covering the MUSE wavelengths. We assume a Moffat profile, with a fixed power index of 2.8, and fit the Full-Width Half Maximum (FWHM) by minimising the difference between a MUSE pseudo F814W image and the \emph{HST} F814W image convolved with the Moffat kernel (see \citealt{Bacon2017} for details).

\subsection{Infrared IFU data}
\label{subsec:sinfoni_data}

MACS1206-arc was targeted with SINFONI in 2011 with a total exposure time of 6 hours. The SINFONI data were reduced with the pipeline developed by MPE (SPRED, \citealt{Abuter2006,Forster-Schreiber2009})  together with custom codes for the correction of detector bad columns, cosmic ray removal, OH line suppression and sky subtraction \citep{Davies2007} and flux calibration.

The main steps of the procedure are as follows. Master bias and
flat images were constructed using calibration cubes taken closest in
time to the science frames and used to correct each data cube. The
science frames were pair-subtracted with an ON-OFF pattern to
eliminate variation in the infra-red sky background. The wavelength
calibration is based on the Ar lamp. For each set of observations, a
flux standard star was observed at approximately the same time and
airmass and was reduced in the same way as the science data. These
flux standard stars were then used for flux calibration by fitting a
black-body spectrum to the O/B stars or a power law to the cold stars
and normalising them to the 2MASS magnitudes. These spectra
were also used to remove atmospheric absorption features from
the science cubes. The different observations were then combined
spatially by using HST images with a larger field of view and good astrometry taken in a similar band as the SINFONI cube, and aligning the SINFONI cube relative to that image. Given that the lenses have such distinctive morphologies, this technique provides reliable coordinates. After these steps, voxels (3D pixels) with flux levels more than 7 standard deviations from the median of the neighbouring voxels were rejected, using a sigma clip algorithm. 

At this point, we inspected the quality of the data. In Fig.~\ref{fig:macs1206_ha} we present the \Ha\, pseudo-narrow band image obtained from the M1206-sys1 data cube by integrating the flux in a spectral window of 12 pixels centred on \Ha\, (which corresponds to 3$\sigma$, assuming a Gaussian shape for the \Ha\, line profile). The continuum was estimated from two close spectral windows of 6 \AA\, width each and subtracted from this pseudo-narrow band. Beside M1206-sys1, other two highly magnified $z\approx1$ galaxies from the sample of \citet{Patricio2018} have been observed with SINFONI: A2390-arc and A521-sys1. However, only M1206-sys1 is bright enough to derive metallicity maps.

Finally, we adjusted the flux calibration and determined the PSF of the SINFONI M1206-sys1 data by comparing a SINFONI F125W pseudo-broad band image with the HST F125W band. The SINFONI field of view is too small to apply the same procedure of image convolution as done with MUSE data, so we fit the two cluster members visible in the SINFONI data. We assume a 2D Moffat profile and, using the {\sc astropy} package \citep{astropy}, fit the cluster members both in HST and in the SINFONI F125W pseudo-broad band image. We then measure the photometry in both images in the same aperture, subtracting the background noise. We find that our nominally reduced SINFONI data overestimate the flux by $\approx11\%$ when compared to HST and we correct the SINFONI data for this offset. We obtained a PSF FWHM of 0.75" and 0.80" for each cluster member, and we take the mean as the seeing of the SINFONI data throughout this work.

\section{Deriving Metallicity}
\label{sec:metallicity}

\begin{figure}
	\includegraphics[width=0.45\textwidth]{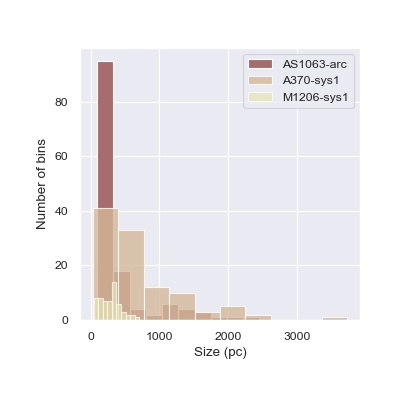}
    \caption{Bin sizes, corrected for magnification, for each galaxy. The sizes were calculated by taking the square-root of the area of each bin.}
    \label{fig:bin_sizes}
\end{figure}

\subsection{Data binning and spectral extraction}
\label{subsec:line_maps}

We start by producing a white light image (the sum along the wavelength axis of the data cubes) for AS1063-arc and A370-sys1 and bin these images using the \citet{Cappellari2003} method of Voronoi tessellation. We opt to use the white light image as opposed to the \oii\, pseudo-narrow band because, in the case of AS1063-arc, using this pseudo-band resulted in a tessellation highly biased towards the strong \hii\, south region. For M1206-sys1, due to the higher noise in the SINFONI cube and the fact that we do not detect significant continuum, we use the \Ha\,pseudo-narrow band. 

We choose a low, arbitrary target signal-to-noise ratio for the tessellation, extract the spectrum from each resulting Voronoi bin, and measure the emission line fluxes in each spectrum (details in the following sub-section). We then check the signal-to-noise ratio of the emission lines in each Voronoi bin. We repeat the process, increasing the target signal-to-noise ratio of the tessellation, until we obtain a signal-to-noise ratio of at least 3 in all bins and for all emission lines. Once this condition is met, we check the quality of the fits of the emission lines for each bin and  reject problematic fits. We use the fluxes measured in each Voronoi bin to derive metallicity, dust attenuation and dust-corrected SFRs.

We check the size of the final bins by summing the number of pixels of each bin and converting this area to physical pc$^2$, using the local value of the magnification to correct for lensing magnification. We then take the square-root of this area as an approximation of the size of the bin and plot this in Fig.~\ref{fig:bin_sizes}. Most of the resulting bins have sizes smaller than 1 kpc.

\begin{figure*}
	\includegraphics[width=\textwidth]{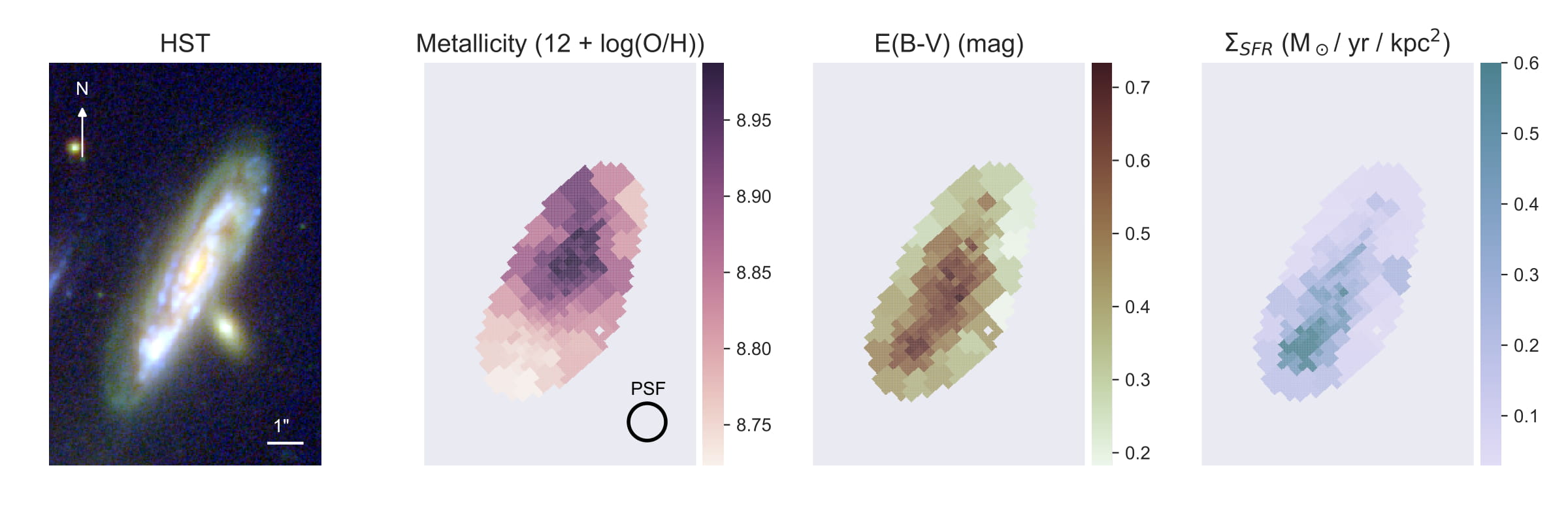}
    \caption{AS1063-arc in the image plane. Left: HST composite image with F160W, F814W and F435W filters. Middle Left: metallicity map. Middle-right: extinction map. Right: SFR surface density map. SFRs were derived from \Hb\, and the \citet{Kennicutt1998} calibration. The FWHM of the PSF is plotted in the lower-left corner of the right panel. All images have the same physical size and orientation.}
    \label{fig:as1063_2Dmet_image_plane}
\end{figure*}

\begin{figure}
    \centering
	\includegraphics[width=0.43\textwidth]{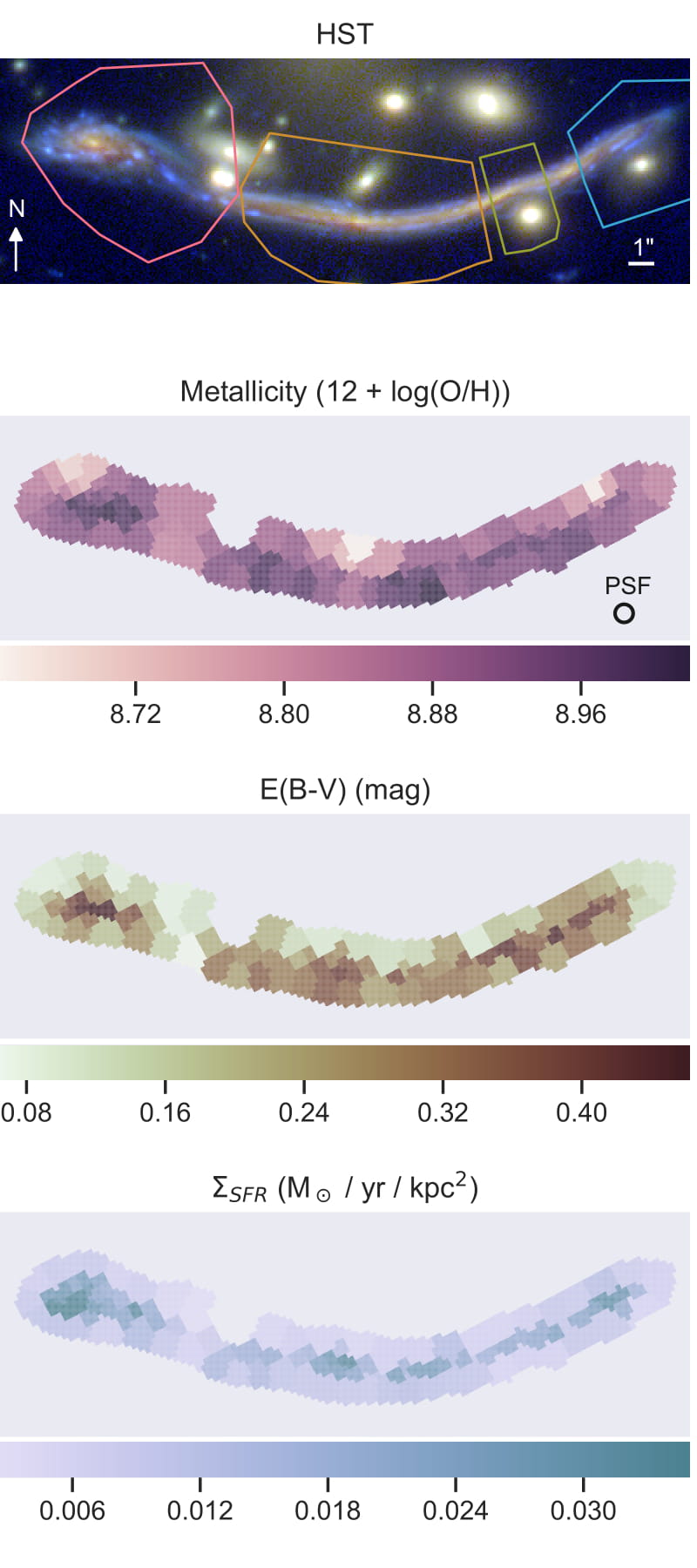}
    \caption{A370-sys1 in the image plane. Top panel: HST composite image with filters F160W, F814W and F435W. Contours correspond to the different multiple images, with the complete image in pink. Lower panels: metallicity, extinction and SFR surface density maps, from top to bottom, each colour coded in a different colour scheme. The FWHM of the PSF plotted in the lower-right corner of the bottom panel. All images have the same physical size and orientation.}
    \label{fig:a370_2Dmet_image_plane}
\end{figure}

\begin{figure}
    \centering
	\includegraphics[width=0.48\textwidth]{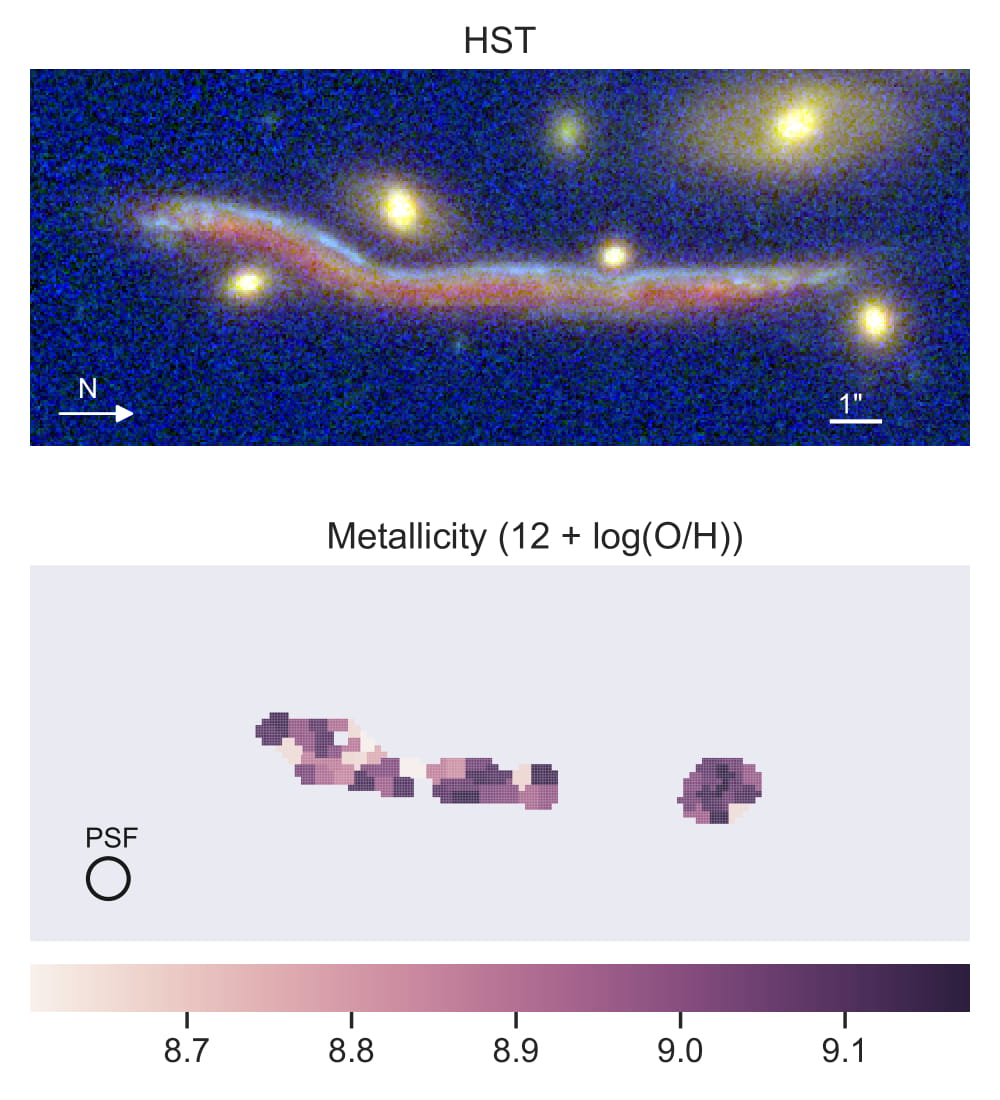}
    \caption{M1206-sys1 in the image plane. Top panel: HST composite image with filters F160W, F814W and F43W5W. Bottom panel: metallicity derived from \Ha\, and \nii. The FWHM of the PSF is plotted in the lower-left corner of the bottom panel.}
    \label{fig:macs1206_2Dmet_image_plane}
\end{figure}

\subsection{Emission line measurements}
\label{subsec:emission_lines_measurements}

The first step in measuring the emission line fluxes is to subtract the continuum, which is especially important for the Balmer lines, since the absorption features are quite significant in these galaxies. We make use of the {\sc pPXF} routine (\citealt{Cappellari2017}, version 6.0.2) and a sample of stellar spectra from the Indo-US library \citep{Valdes2004}. The continuum fit is performed masking emission lines. To improve the fit, we add a low-order polynomial to the templates and multiply by a first order polynomial. 

After this, the continuum is subtracted from the spectrum and the emission lines are measured using the Automated Line Fitting Algorithm ({\sc alfa}) from \citet{Wesson2016}. Comparing results obtained using {\sc alfa} and the method of \cite{Patricio2018}, we obtained flux differences of less then 8\% in the integrated spectra for fainter Balmer lines (\Hd\, and H7) and less than 1\% for strong emission as \oii.

We present the resulting emission line maps, as well as the maps of the ratios used to derive metallicity in the following subsection, in Appendix~\ref{app:line_maps}.

\subsection{Determining metallicity, SFR surface density, and extinction}
\label{subsec:method}

We use the following diagnostics to derive the metallicity in our sample:

\begin{description}[align=right,labelwidth=1cm]
\item [\emph{O2}] \oii/\Hb
\item [\emph{O3}] \oiiib/\Hb
\item [\emph{O32}] \oii/\oiiib 
\item [\emph{R23}] (\oii+\oiiia+\oiiib)/\Hb
\item [\emph{N2}] \nii/\Ha
\end{description}

with the \emph{O2, O3, O32, R23} diagnostics and \Hb/\Hg\, being used for AS1063-arc and A370-sys1, and \emph{N2} for M1206-sys1. We use the \cite{Maiolino2008} strong line calibration to derive metallicities from these line ratios. Since \Hb\, is not available for M1206-sys1, the \emph{O2} diagnostic was derived by extrapolating the \Hb\, flux using the intrinsic \Hg\, (i.e. dust corrected, see below) from the MUSE data, assuming the \Hb/\Hg\, ratio of 2.135, for T$_e=10000$ K and low electron density and case B recombination \citep{Storey1995}. 

We make use of the \Hb/\Hg\, ratio to derive the redenning correction in the case of AS1063-arc and A370-sys1. For M1206-sys1, no correction is applied to the \emph{N2} ratio, due to the large uncertainties when combining MUSE (\Hb,\,\Hg) with SINFONI data (\Ha). Moreover, the proximity of \Ha\, and \nii\, makes the differential dust attenuation between these two lines small enough that it is reliable to derive metallicities without including dust-correction. 

We do not correct for Galactic extinction. This correction would be very small in the case of AS1063-arc and A370-sys1 ($E(B-V)$ = 0.012 mag and 0.032 mag respectively), and with a variation of less 0.001 mag within the full length of the gravitational arcs \citep{Schlafly2011}. For M1206-sys1, the galactic extinction is higher ($E(B-V)$ = 0.063 mag), but for the reason mentioned above, we do not apply this correction either.

Finally, we derive metallicity ($Z$) and attenuation (E(B-V)) from several emission line ratios ($r$) in a Bayesian framework, fitting multiple strong line metallicity diagnostics and extinction simultaneously. We use the {\sc emcee} Markov chain Monte Carlo Sampler \citep{Foreman-Mackey2013} to maximise the following Gaussian ($\log$-)likelihood function:

\begin{eqnarray}
\label{eq:lnprob}
ln\,p = - \frac{1}{2} \sum_r \left[ \left(\frac{{\rm M_r}(Z) - {\rm O_r}(E(B-V))}{\sigma_r^2}\right)^2 + \ln(2\pi\sigma_r^2) \right]
\end{eqnarray}

where ${\rm O_r}(E(B-V))$ are the observed line ratios corrected for attenuation and ${\rm M_r}(Z)$ are the respective expected ratios, obtained from the \cite{Maiolino2008} calibrations. $\sigma_r^2$ is the quadratic sum of the observed error and an additional model uncertainties. We adopt an uncertainty of 0.1 dex for the metallicity calibrations and a 1\% uncertainty for the case B Balmer line ratios. We use a wide flat prior on metallicity, between $7.0 <\met<9.2$, the range of the data analysed in \cite{Maiolino2008} (see their figure 5), and a wide flat prior on attenuation of $0<$E(B-V)$<1$ mag. 

The star-formation rates densities are calculated by taking the \Hb\, intrinsic fluxes and calculating the expected intrinsic \Ha\, fluxes, assuming case B, a temperature of T=10 000K and low electron density, and applying the \cite{Kennicutt1998} calibration. Since we calculate SFR densities, no magnification corrections are needed because gravitational lensing conserves surface density brightness (the increased flux due to lensing covers a larger area). The dust attenuation also does not depend on lensing correction, since it is derived from line ratios of each pixel.

We adopt this Bayesian approach as a systematic way to combine different indicators, which has the advantage of having a self-consistent dust and metallicity treatment. However, we do not claim that this will necessarily yield statistically meaningful uncertainty estimates, since the line ratios used in the likelihood function are not independent from each other.

In order to estimate uncertainties in an alternative way, we compute the metallicity using each diagnostic  independently and calculate the dispersion of values obtained for each bin (see Appendix~\ref{app:line_maps}). We did not include any dust correction in these calculations. For AS1063-arc, we obtain a mean  standard deviation between metallicity values of 0.09 dex and a maximum dispersion of 0.24 dex, compared with a mean and maximum of 0.03 dex and 0.04 dex obtained using our Bayesian approach. For A370-sys1, we obtain a mean and maximum dispersion of 0.05 dex and 0.12 dex from the individual diganostics, compared with 0.03 dex and 0.07 dex from the Bayesian likelihood maximisation. We notice that amongst the four diagnostics included -- \emph{R23, O3, O2} and \emph{O32} -- the latter is the one that most deviates from the mean for both galaxies.

Since the \emph{O32} ratio is sensitive to the ionisation parameter \citep[e.g.][]{Kewley2002}, it is possible that differences in local ionisation parameter are driving the dispersion in metallicity. For AS1063-arc, this diagnostic deviates most from the metallicities calculated with the other 3 diagnostics in the \hii\, south region, where the highest SFR densities are also found (see Fig.~\ref{fig:as1063_2Dmet_image_plane}) and the highest ionisation parameters is expected due to recent star-formation, which seems to further confirm this hypothesis. It is worth noticing however that the relation between SFR and ionisation parameter is not fully established. For example, \cite{Paalvast2018} do not find a relation between sSFR and the O32 ratio. Furthermore, \cite{Shirazi2014} suggest that high-z galaxies with elevated O32 ratios have high electron densities, not necessarily higher ionisation parameters.

\subsection{Metallicity Maps}
\label{subsec:met_maps}

For both AS1063-arc and A370-sys1, \oii, \oiiia, \oiiib, \Hg, and \Hb\, can be measured with a signal-to-noise of at least 3 in each bin. We derive the metallicity and extinction maps using a total of 6 line ratios (\emph{O2, O3, O32, R23}, \mbox{[O\,{\scshape iii]}}\,$\lambda$5007/4959, \Hb/\Hg). Weaker lines, such as \neiii\, and H7 are also well detected in the integrated spectra of these galaxies, but cannot be used to derive resolved properties due to their faintness. We present the comparison between the metallicities derived using different line sets, with and without these fainter lines, in Appendix~\ref{app:integrated_metallicity}. The resolved maps of metallicity, SFR and extinction for these two galaxies are shown in Fig.~\ref{fig:as1063_2Dmet_image_plane} and \ref{fig:a370_2Dmet_image_plane}.

For M1206-sys1, we derive the metallicity using the \nii/\Ha\, ratio and not do include dust-correction. We show the metallicity map of M1206-sys1 in Fig.~\ref{fig:macs1206_2Dmet_image_plane}.

We did not account for Diffuse Ionised Gas (DIG) in this analysis, which might impact the values of metallicity. In their sample of local disc galaxies with resolutions of $\approx100$ pc, \cite{Erroz-Ferrer2019} found that the DIG regions have metallicity on average 0.1 dex lower than the \hii\, regions, so we might assume our metallicity values to be upper limits. On the other hand, \cite{Erroz-Ferrer2019} found that the radial gradient of both metallicities (\hii\, regions and DIG) were similar, so this caveat in our analysis might not impact the derived gradients, if this result is also valid at $z\approx1$.

\section{Resolved Scaling Relations}
\label{sec:local_fmz}

We start our analysis by checking whether the resolved Star-Forming Main Sequence (rSFMS) and the resolved Mass-Metallicity Relation (rMZR) are in place for these galaxies. Since these relations only involve surface density quantities (or metallicity) which are conserved by gravitational lensing, we can investigate these correlations regardless of lensing correction.

\subsection{Resolved Mass-SFR relation}

We derive stellar mass surface densities ($\Sigma_{\star}$) by measuring the photometry in multiple HST bands (F105W, F110W, F125W, F140W, F160W, F435W, F606W, F625W, F775W, F814W and F850W) for each bin defined in the MUSE data. We then use {\sc FAST}\footnote{We included both spectra and photometry to derive the total mass using {\sc prospector} in our previous work. Since in this case we only use photometry (the spectral continuum signal-to-noise ratio is too low to further constrain the fit), we opted to use {\sc} FAST, since the computational time to derive masses is substantially smaller.} \citep{Kriek2009}, with the \cite{Bruzual2003} stellar synthesis models, the \cite{Chabrier2003} IMF and an exponentially decaying star-forming history, and a \citet{Calzetti2000} dust attenuation law. We convert the output masses into mass surface densities, which, as stated before, is independent of lensing.

\begin{figure*}
	\includegraphics[width=0.9\textwidth]{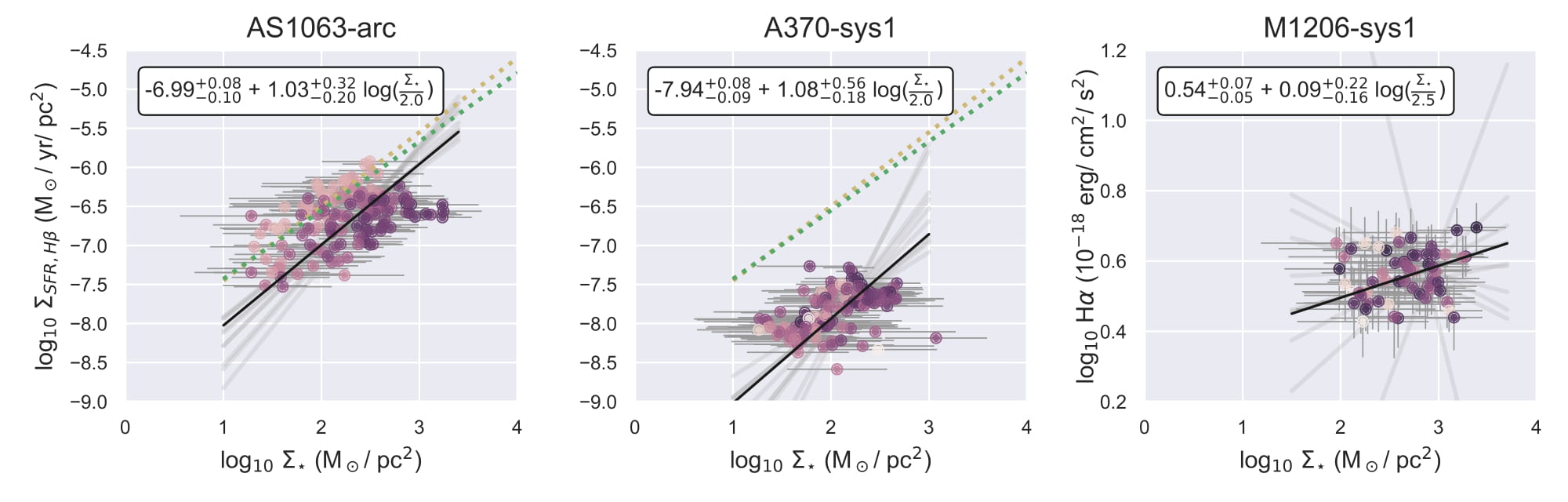}
	\includegraphics[width=0.9\textwidth]{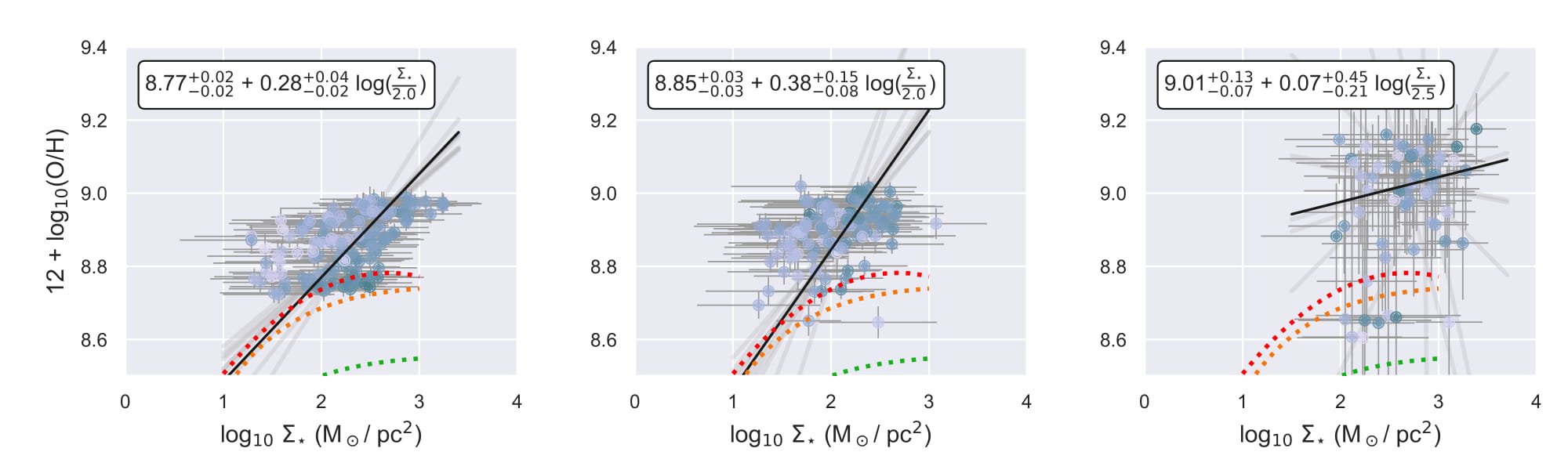}
	\includegraphics[width=0.9\textwidth]{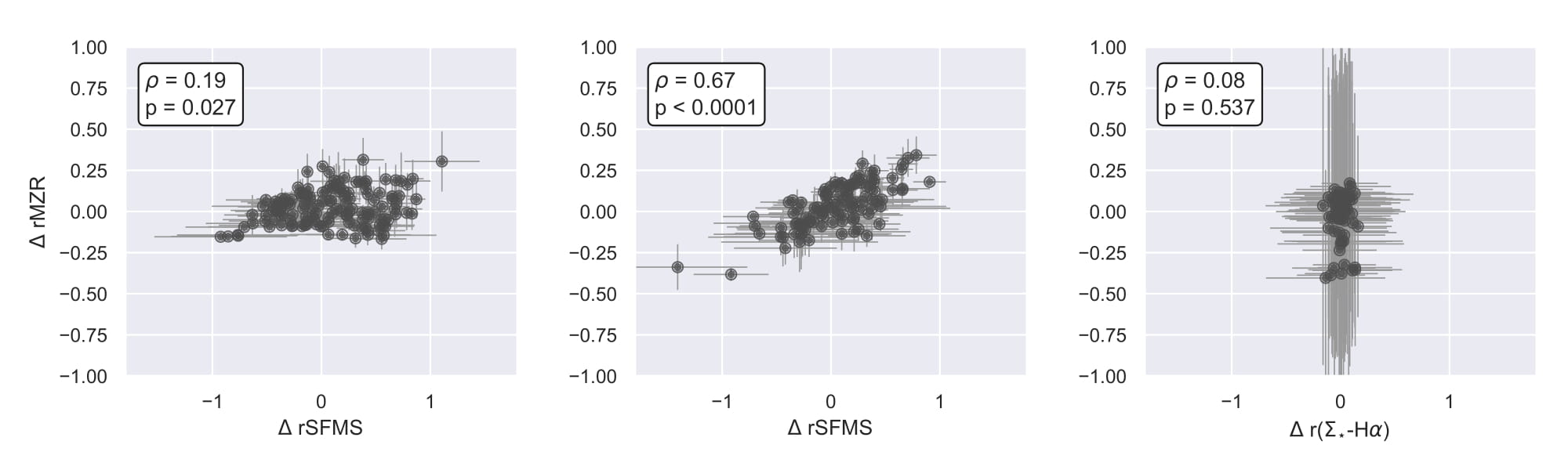}
    \caption{Resolved scaling relations. Top row: rSFMS, colour-coded by metallicity, with higher metallicities in darker colours. For M1206-sys1, since we can not derive the resolved dust correction, we report \Ha\, fluxes instead of $\Sigma_{SFR}$. We plot the results from \citet{Wuyts2013} in yellow and the results from \citet{Abdurrouf2018} in green. Middle row: rMZR, colour-codded by SFR. We also plot the results of \citet{Rosales-Ortega2012} in red, \citet{Sanchez2013} in orange, and \citet{Barrera-Ballesteros2016} in green. The linear fit results are plotted in the upper-left corner of each plot and possible realisations of this fit are plotted in grey lines. Bottom row: residuals of the rMRZ versus residuals of rSFMS (and the r$\Sigma_{\star}$-$H\alpha$ residuals for M1206-sys1). Uncertainties were calculated including the linear fit uncertainties. We show the Spearman correlation rank ($\rho$) and the $p$ value of these correlations in the upper-left corner.}
    \label{fig:scaling_relations}
\end{figure*}

Using these mass densities and the star-formation rate densities derived from the H$\beta$ lines ($\Sigma_{SFR, H\beta}$) for AS1063-arc and A370-sys1, we plot the rSFMS in the first row of Fig.~\ref{fig:scaling_relations}. For MACS1206-sys1, we use the flux of H$\alpha$ as a proxy for SFR, although this is merely indicative.

We fit the rSFMS using a hierarchical Bayesian model, {\sc linmix}\footnote{https://linmix.readthedocs.io} \citep{Kelly2007}, that fits a linear model taking into account uncertainties in both variables involved in the relation. We fit a linear model in the form $\log_{10}\Sigma_{SFR} = a + b\, \log_{10} \left(\frac{\Sigma_\star}{2.0}\right)$ for AS1063-arc and A370-sys1 and $\log_{10}\Ha = a + b\, \log_{10} \left(\frac{\Sigma_\star}{2.5}\right)$ for M1206-sys1, placing the pivot point of the linear relation in the middle of the data. We plot the resulting fits in Fig.~\ref{fig:scaling_relations}. We obtain slopes of b = 1.03$^{+0.32}_{-0.20}$ and 1.08$^{+0.56}_{-0.18}$ for AS1063-arc and A370-sys1, confirming that the SFMS is locally present in these two galaxies. These uncertainties were calculated by taking values of the slope and intercept from several steps of the {\sc limix} MCMC chain, and placing them in histograms. Since some of the resulting distributions are asymmetric, we take the upper and lower errors as the minimum and maximum values that contain 68\% of the values centred in the maximum of the histogram. For a Gaussian distribution, this corresponds to the 1$\sigma$ error.

These slopes agree, within uncertainties, with what was obtained by \cite{Wuyts2013} using 473 massive star-forming galaxies at $0.7<z<1.5$ at kilo-parsec resolutions (slope of 0.95, in yellow dotted line in Fig.~\ref{fig:scaling_relations}). In a recent work, \cite{Abdurrouf2018} also analysed the rSFMS at 1 kpc resolution for massive disc galaxies at $0.8<z<1.8$ (slope of 0.88, in green dotted line), calculating SFRs from broad band SED fitting, finding similar results to \cite{Wuyts2013} and the ones derived here.

For M1206-sys1, there seems to be no correlation between the mass density and the H$\alpha$ flux (the slope is compatible with zero), which might be an indication that the dust attenuation is not the same in the entire galaxy.

\subsection{Resolved Mass-Metallicity relation}

We plot the metallicity derived for each bin and the corresponding stellar surface density masses in the middle row of Fig.~\ref{fig:scaling_relations} in order to study the rMZR. For AS1063-arc and A370-sys1, we find that metallicity and stellar mass density are correlated, with higher density bins having higher metallicities. Although at lower redshifts (and with substantial more data) this relation is fit with a more complex function, given the small range explored by our data (2 orders of magnitude in $\Sigma_\star$), we fit the relation with a linear model, as done for the rSFMS.

We obtain different slopes of 0.28$^{+0.04}_{-0.02}$, 0.38$^{+0.15}_{-0.08}$ for AS1063-arc and A370-sys1, which are compatible within uncertainties. For M1206-sys1, we obtain a slope of 0.07$^{+0.45}_{-0.21}$ between metallicity and stellar mass density, which indicates a very weak relation between these two quantities.

We also plot the relations obtained in the local Universe using the PINGS \citep{Rosales-Ortega2012} in red, CALIFA \citep{Sanchez2013} in orange (we use their equation (1) with the parameters a = 8.74, b = 0.018, c = 3.05 (S\'anchez, private com., also used in \cite{Barrera-Ballesteros2016}), and MaNGA \citep{Barrera-Ballesteros2016} in green, using more complex functional forms to fit this relation. Our data points generally fall above all these local relations, i.e. they all have higher metallicities for the same mass surface density than what is found in the local Universe.

However, since determining absolute calibrations is still challenging, it is difficult to directly compare results obtained in different works. Both \cite{Rosales-Ortega2012} and \cite{Sanchez2013} use the \emph{O3N2} ratio ((\oiiia/\Hb) / (\nii/\Ha)) and the calibrations of \cite{Pettini2004} (PP04), while \cite{Barrera-Ballesteros2016} use the same ratio but with the \cite{Marino2013} (O3N2-M13) calibrations. \cite{Sanchez2017} investigate these differences calculating the MRZ using different metallicity calibrators for a sample of 613 galaxies observed in the CALIFA survey, obtaining for the same mass, differences of up to 0.4 dex between calibrations. In this analysis, the \emph{O3N2} calibrations of \cite{Pettini2004} and \cite{Marino2013} are included as well as the \emph{R23} from \cite{Maiolino2008} (M08), that we will take as a good approximation to the results derived here combining \emph{R23, O3, O2} and \emph{O32}.

The O3N2-M13 calibration gives results up to 0.2 dex lower than the ones with M18 (figure 3 of \citet{Sanchez2017}), which might explain why the rMZR of \cite{Rosales-Ortega2012} (in red in Fig.~\ref{fig:scaling_relations}) predict lower metallicities for the mass-densities analysed here (see Fig.~\ref{fig:scaling_relations}). However, the PP04 calibrations give similar results as the M08 calibrations ($\approx0.02$ dex), while the results from \cite{Barrera-Ballesteros2016} (in orange in Fig.~\ref{fig:scaling_relations}) are the ones that most deviate from our results.

It is then difficult to say with certainty if there is an evolution with redshift of the rMRZ or if the discrepancies seen here arise due to the differences in metallicity calibrations.

\cite{Trayford2018} used the EAGLE simulation to study the evolution of the rMZR with redshift. They find a strong evolution in the shape of this relation when AGN feedback is included, while it remains fairly similar from $z=0.1$ to 2 when no AGN are present. However, even in this last case, the normalisation (i.e. intercept) of the rMZR shows a strong evolution of about 0.4 dex for stellar mass densities of 10$^2$ M$_\star$/pc$^{2}$, with higher-$z$ having lower metallicity values.

For the same range of stellar-mass densities studied here, we find metallicity values that are $\approx0.4-0.5$ dex higher than predicted by \cite{Trayford2018} for $z=0.5$ and 1. As for the observational studies, it is not clear if this difference is driven by the choice of metallicity calibration.

\subsection{Resolved Fundamental Mass-Metallicity relation}

Finally, we investigate the correlation between the residuals of the rSFMS and rMRZ. We plot this in the lower panel of Fig.~\ref{fig:scaling_relations} and calculate the Spearman correlation test for these two residuals. 

For AS1063-arc we measure a correlation of $\rho=0.19$ (with corresponding $p$ value of 0.027), corresponding to a weak correlation. For A370-sys1 we obtain a strong correlation of $\rho=0.67$ ($p<$0.0001). For M1206-sys we compare the residuals of the rMZR with the ones from the stellar mass density vs \Ha\, flux, that we denoted as r($\Sigma_\star$-\Ha), and found no clear correlation between these residuals ($p=0.537$).

Excluding M1206-sys1 from the analysis, given our lack of $\Sigma_{SFR}$ for this galaxy, we measure a positive correlation between $\Delta$rSFMS and $\Delta$rMZR for AS1063-arc and A370-sys1, although weak in the case of AS1063-arc. This might indicate that a relation between resolved $\Sigma_{\star}$, $\Sigma_{SFR}$ and Z is present at higher$-z$. However, given the different values we obtained for this correlation in these two galaxies of comparable mass and metallicities, it might indicate that this relation is not fundamental, in the sense that it is not the same for all galaxies at all redshifts. 

We notice also that we find a positive correlation between the two residuals, with higher residuals in rSFMS corresponding to higher residuals in rMRZ, instead of the negative correlation between SFRs and metallicity, for fixed stellar mass, found in other works \citep[e.g.][]{Lara2010,Mannucci2010}, with higher residuals in rSFMS corresponding to higher residuals in rMRZ. However, we base these conclusions in only two objects, and a larger sample with wider redshift range is needed in order to confirm these results.

\begin{table*}
\caption{AS1063-arc and A370-sys1 metallicity gradient and morphology fit. GALFIT: results of the morphological fit to the reconstructed F160W HST band using {\sc galfit}. Remaining rows: fit of the image plane metallicity gradient using the procedure described in \ref{subsec:method}, fixing or letting the morphological parameters vary.}
\label{tab:frapy_fit}
\centering
\tabcolsep=0.1cm
\begin{tabular}{|lcccccccc|} 
\hline
\multicolumn{9}{c}{AS1063-arc}\\
\hline
         &  & $\nabla$Z   & Z$_{0}$ & Centre RA  & Centre Dec  & q &$\theta$  & $\chi^{2}$/dof \\
         &  & [dex/kpc]   & [\met] & J2000  & J2000  &  &[deg]  &  \\
\hline
{\sc galfit} & -  & -   & -   & 22h48m42.859s  & -44d31m57.0464s  & 0.56 & -32    &    37.25          \\
Fixed morph. & prior    &  [-0.1:0.0] & [8.5:9.5]  & 22h48m42.859s   & -44d31m57.0464s  & 0.56 & -32   &   \\
                         & fit  &  -0.034$\pm$0.002  &   8.985$\pm$0.007 & -  & -  & - & -  & 6.04 \\
Free par.    & prior    &  [-0.1:0.0] & [8.5:9.5]  & 22h48m[41.634\,:\,41.871]s   & -44d31m[55.169\,:\,57.708]s  & [0.1:0.9] & [-90:90]   &   \\
                        & fit  & -0.042$\pm$0.002  &  9.038$\pm$0.008 & 22h48m41.750s & -44d31m56.016s  & 0.52$\pm$0.05 & 68$\pm$2  & 1.44 \\
\hline
\multicolumn{9}{c}{A370-sys1}\\
\hline
         &  & $\nabla$Z   & Z$_{0}$ & Centre RA  & Centre Dec  & q &$\theta$  & $\chi^{2}$/dof \\
                  &  & [dex/kpc]   & [\met] & J2000  & J2000 &  &[deg]  &  \\
\hline
{\sc galfit} & -    & -   & -   & 02h39m53.716s   & -01d35m03.55s  & 0.32 & -52      &  42.04  \\
Fixed morph. & prior    &  [-0.1:0.0] & [8.5:9.5]  & 02h39m53.716s   & -01d35m03.55s  & 0.32 & -52   &   \\
 & fit  & -0.039$\pm$0.004  &   8.980$\pm$0.007 & -  & -  & - & -  & 4.79 \\
Free par. & prior    &  [-0.1:0.0] & [8.5:9.5]  & 02h39m[53.573\,:\,53.805]s   & -01d35m[02.921\,:\,07.817]s  & [0.1:0.9] & [-90:90]   &   \\
& fit  &  -0.053$\pm$0.004 &  9.032$\pm$0.009 & 02h39m53.709s & -01d35m04.169s  & 0.39$\pm$0.04 & -47$\pm$3  & 3.80 \\
\hline
\end{tabular}
\end{table*}

\section{Metallicity Gradient}
\label{sec:fit}

We now turn our attention to the metallicity distribution within each galaxy, deriving its gradient and inspecting the residuals. We start by describing how gravitational lensing affects the galaxy properties, and proceed to describe how we model the data with a simple 2D radial gradient, taking into account lensing and seeing effects with a forward-modelling approach.

\subsection{Lensing distortion}
\label{subsec:lensed_morphology}

AS1063-arc is the least magnified galaxy, with a mean magnification of $\mu=4$, and also only a small distortion. Using {\sc Lenstool} \citep{Jullo2007} and the respective lensing model, we can reconstruct the morphology in \textit{source plane}, i.e. corrected for lensing magnification (see Fig~\ref{fig:as1063_source_plane}). This process does not account for seeing effects, and the PSF in the source plane is not circular, with a smaller FWHM in the direction where the galaxy is more magnified, where effectively we can probe smaller spatial scales (see the second panel in Fig~\ref{fig:as1063_source_plane}, in appendix). This means that spatial resolution is not homogeneous in this galaxy, which we will explore in the next section. 
 
A370-sys1 and M1206-sys1 have higher magnification factors, reaching $\mu=30$ in some regions, and more complex lensed morphologies, with multiple images of the same regions, which makes the reconstruction process more challenging. The lensed image of A370-sys1 contains one complete image of the galaxy, plus 3 other partial images, i.e., only a portion of the galaxy was imaged into those multiple images. This is also the case for M1206-sys1, where 4 multiple images can be seen in the SINFONI data. However, unlike A370-sys1, the SINFONI data do not contain the full image, and only about half of the disc is available. 

Each of these multiple images can be traced back to the source plane using the lensing models. However, this leads to different PSFs in the source plane, since their lensing distortions are different. For AS1063-arc, the FWHM of the PSF measures 2.3 kpc in the direction of highest magnification and 5.69 kpc in the lowest and for A370-sys1, between 0.73 and 3.10 kpc. This means that combining several multiple images in the source plane, without including seeing deconvolution, can produce misleading results. Strategies to deal with this issue have been developed \citep{Sharma2018}, but here we choose a simpler approach, and perform most of our analysis in the image plane, keeping the multiple images separated. 

\subsection{Forward-modelling metallicity gradients}
\label{subsec:2dfit}

In order to fully use the spatial information provided by the IFU observations, we fit the metallicity maps assuming a simple 2D axisymmetric gradient, where the metallicity depends on the deprojected galactocentric distance to the centre of the galaxy (corrected for inclination and lensing), the assumed gradient ($\nabla Z$) and central metallicity value ($Z_0$).

We build our gradient model in the source plane, calculate the lensing distortions using the lensing models, convolve the lensed gradient with the instrument seeing, and finally compare it with the data, minimising the difference between the two. This approach is similar to the one presented in \citet{Carton2017} for field galaxies, but includes the lensing correction. 

We start by producing a deprojected galactocentric distance 2D map in the source plane, using the centre of the galaxy (c$_{x}$, c$_{y}$), the ratio between the minor and the major axis (q), and the position angle ($\theta$). Using the lensing model, we forward-lense this deprojected galactocentric distance map to the image plane and align it with the data, rescaling the pixel sizes to match the IFU observations. We then multiply this map by the gradient and add the central metallicity value (Z(x,y) = Z$_0$ + $\nabla$\,Z\,r) to produce a metallicity gradient in the source plane. Finally, we convolve it with the seeing, and apply the same binning as used to derive the metallicity maps. We compare this model gradient with the measured metallicity maps using a Gaussian log-likelihood function and the {\sc emcee} sampler to maximise the likelihood and obtain the best-fit parameters (as done Section~\ref{subsec:method}). We have made this method publicly available\footnote{The code, {\sc FRApy}, for Fitting Resolved Arcs with {\sc Python}, is available at \url{https://frapy.readthedocs.io}.}.

\begin{figure}
	\includegraphics[width=0.23\textwidth]{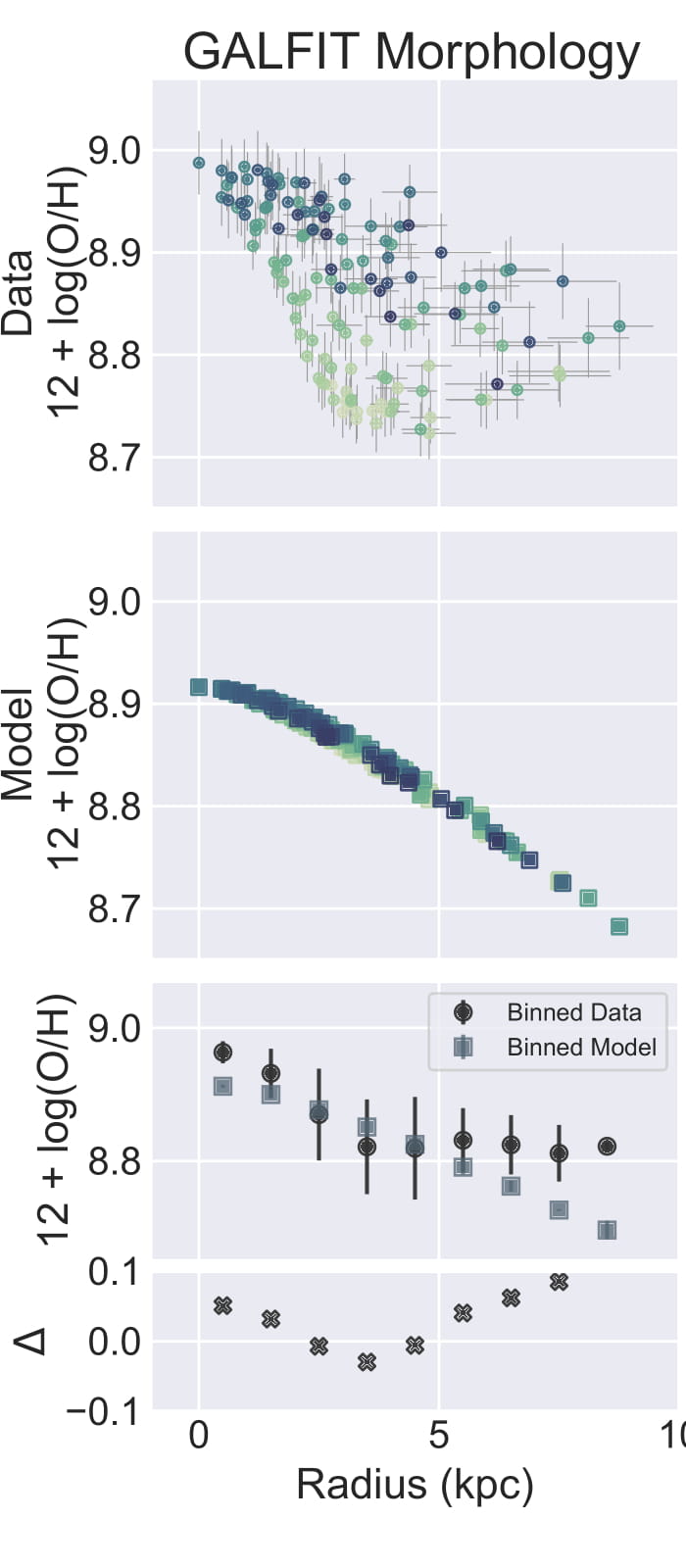}
	\includegraphics[width=0.23\textwidth]{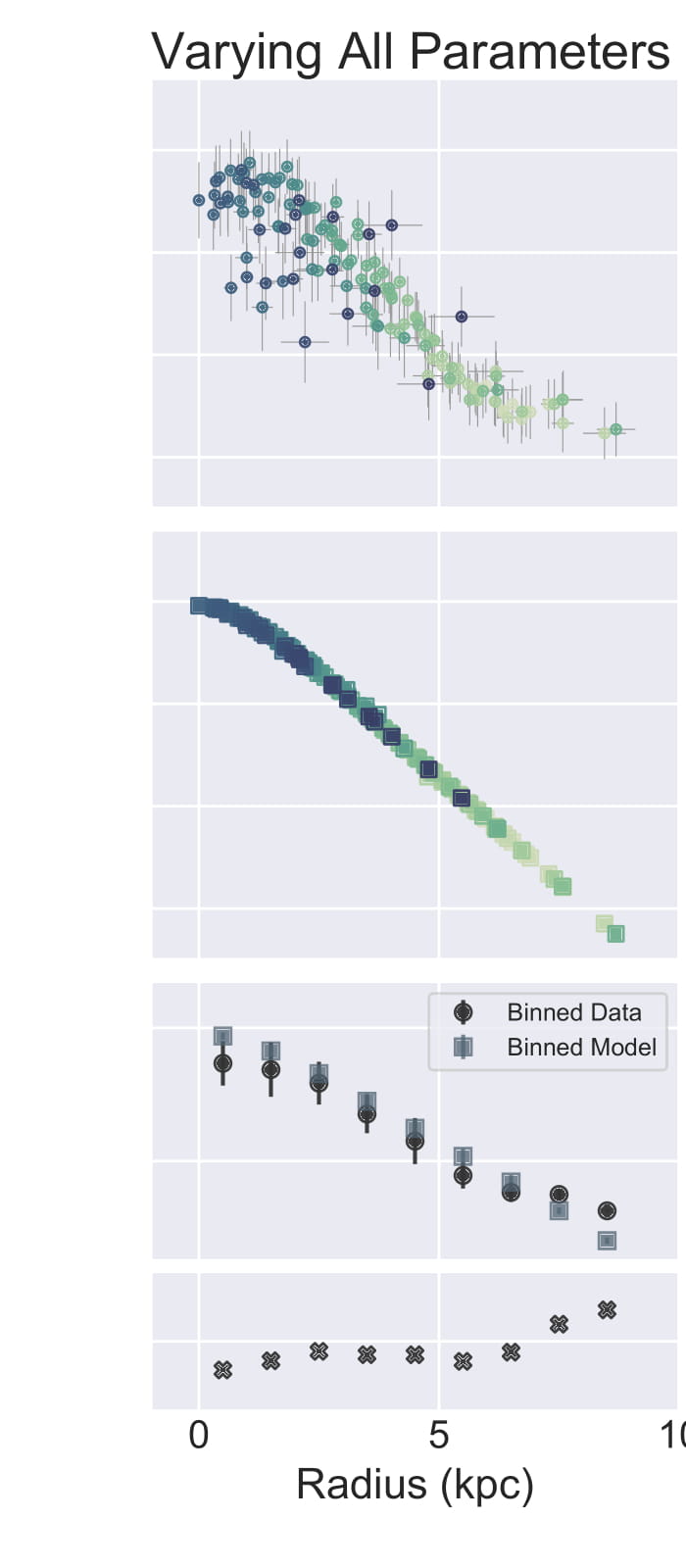}
    \caption{AS1063-arc radial variation of metallicity using the morphology derived from HST with {\sc galfit} (left panels) and letting the morphological parameters free (right panels). Data is shown in the top panels and the model gradient convolved with the seeing is shown in the middle panel. Each point corresponds to a Voronoi bin, colour coded by the number of the bin, so that the same bin has the same colour in all plots and adjacent bins have similar colours. The lower panels display the binned version of both the data (circles) and the model (squares) and the binned residuals (crosses).}
    \label{fig:as1063_1d}
\end{figure}

\begin{figure}
	\includegraphics[width=0.23\textwidth]{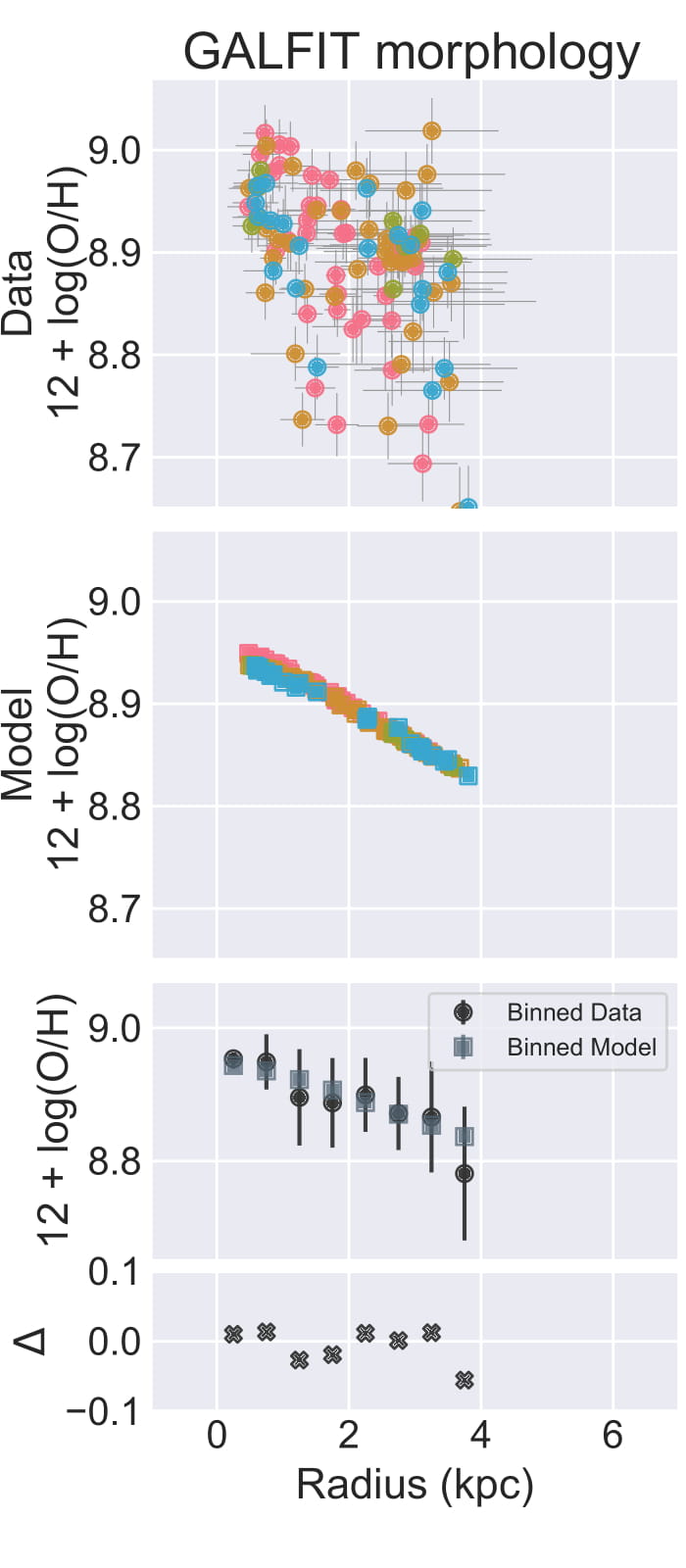}  
	\includegraphics[width=0.23\textwidth]{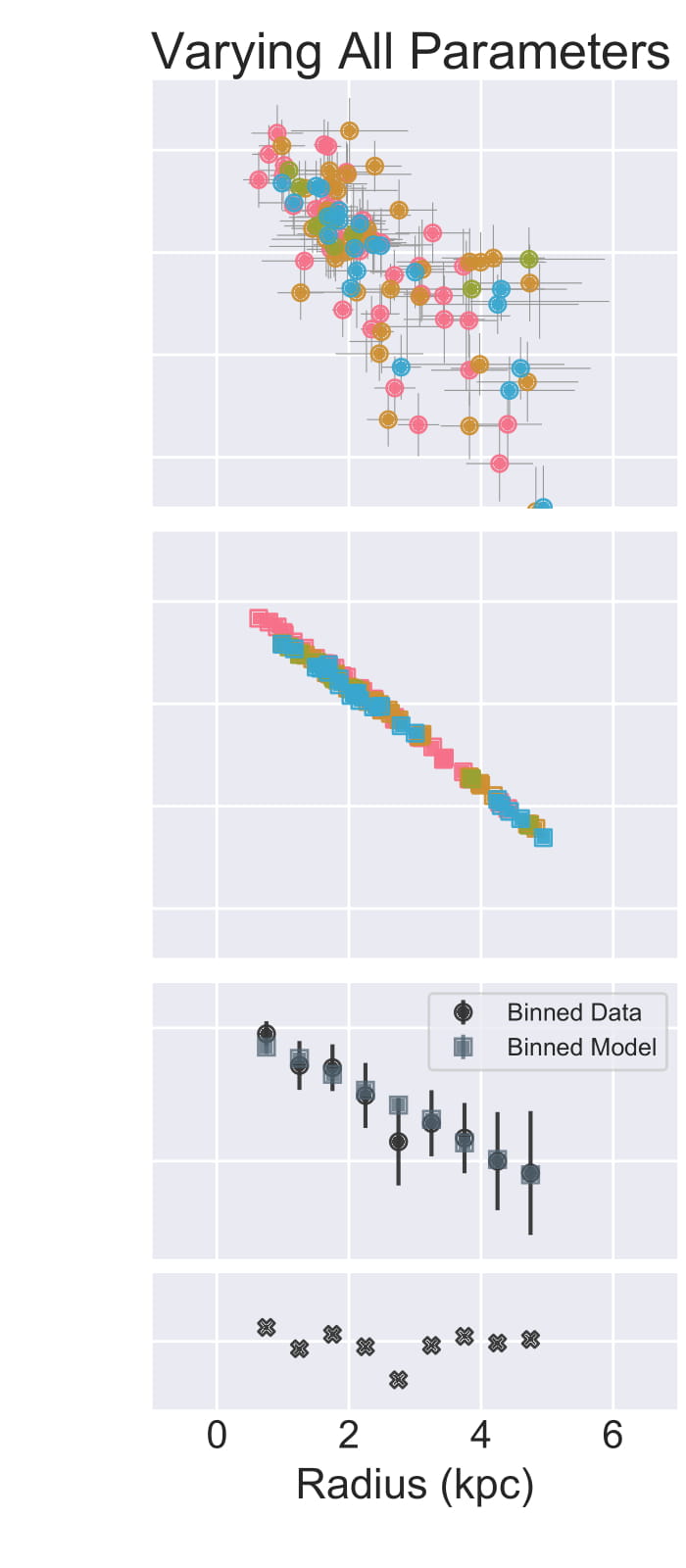}
	\caption{As Fig.~\ref{fig:as1063_1d} but for A370-sys1.}
    \label{fig:a370_1d}
\end{figure}

\subsubsection{AS1063-arc}

We start by producing a source plane image of the F160W HST band and fit it with {\sc galfit} \citep{Peng2010}, in order to assess what values the morphological parameters of the metallicity gradient model -- q, $\theta$ and centre -- could have. We used a global S\'ersic profile plus two more compact components for the bulge and the large \hii\, southern region. We report the relevant results of the fit in Table~\ref{tab:frapy_fit}. 

We fit the data first keeping q, $\theta$ and the centre fixed to the values obtained with {\sc galfit}, and then letting them vary within large intervals. The morphological parameters obtained in the second case are very different from what was obtained with {\sc galfit}. The centre is offset by about 0.4 arcsec and the position angle $\theta$ differs by $\approx90$ degrees. This difference arises from the fact that the two spiral arms (and the major axis of the galaxy derived with {\sc galfit}) are aligned with the direction of the highest stretch caused by gravitational lensing, that together with the poor seeing at which this galaxy was observed ($\approx1$"), makes it challenging to derive the correct morphology. 

Following these two approaches, we obtain gradients of -0.034$\pm$0.002 and  -0.042$\pm$0.002 dex/kpc, respectively and central metallicities (8.99$\pm$0.01 and 9.04$\pm$0.01 in 12+log(O/H)). We plot the 1D profiles for both these fits in Fig.~\ref{fig:as1063_1d}.

\subsubsection{A370-sys1}

We fit the A370-sys1 metallicity map with the same technique, starting by fitting the morphology using the F160W HST band. Due to the difficulties in combining different multiple images (see subsection ~\ref{subsec:lensed_morphology}), we use only the complete multiple image to perform the {\sc galfit} fit. Since this galaxy also has a complex morphology, we use several components in the fit (disc, bulge plus strong star-forming regions), and report the values for the disc in Table~\ref{tab:frapy_fit}.

We then proceed to fit A370-sys1 fixing the morphology to the values found with {\sc galfit} and also letting q, $\theta$ and the central position free. The results are listed in Table~\ref{tab:frapy_fit}. In this case, we obtain axis ratios and $\theta$ closer to what was obtained with {\sc galfit}, but still inconsistent with this method.

The central metallicities obtained in both fits are also close (8.98$\pm$0.01 and 9.03$\pm$0.01 in 12+log(O/H)), and although not formally compatible, they are well within the typical uncertainty of metallicity calibrations. We also obtain different gradients, -0.039$\pm$0.004 and -0.053$\pm$0.004 dex/kpc, respectively. The 1D profiles obtained with both fits are shown in Fig.~\ref{fig:a370_1d}.

\subsubsection{M1206-sys1}

Because of the complexity of the lens model and the low(er) number of metallicity measurements, which do not allow us to reliably constrain the parameters of the metallicity gradient model, we performed only a simple 1D analysis for M1206-sys1. 

\begin{figure}
	\includegraphics[width=0.48\textwidth]{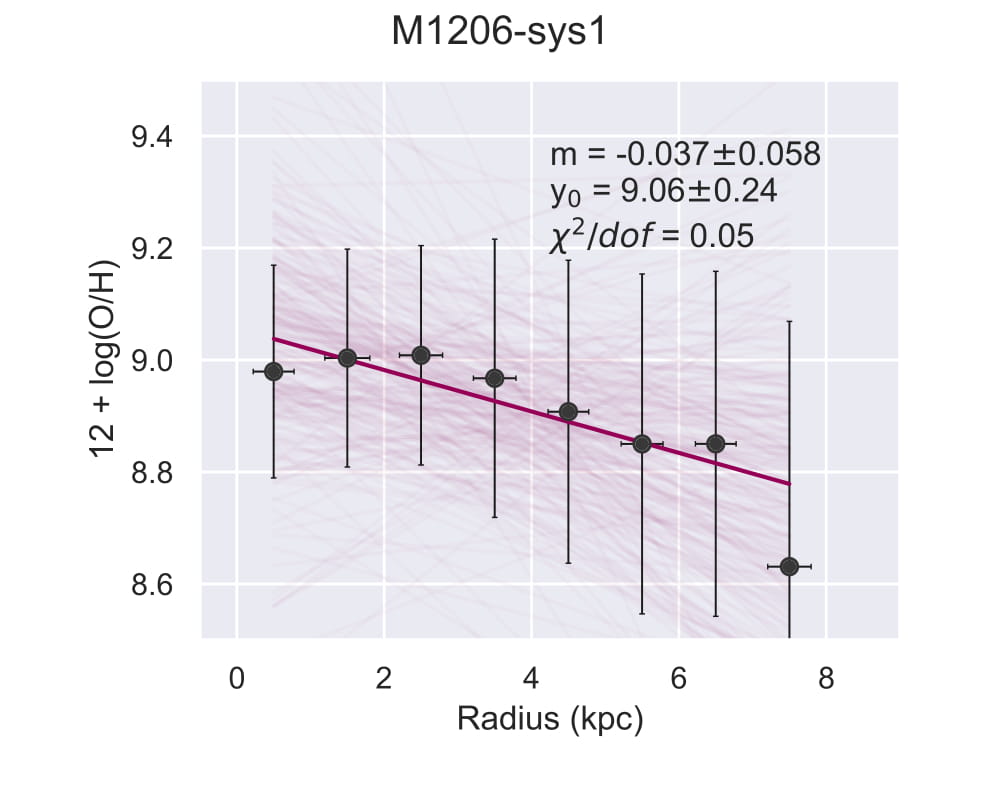}
    \caption{M1206-sys1 radial variation of metallicity. The data points correspond to averages within annuli. The fit was performed with the {\sc linmix} package. The pink lines are multiple realisations of the fit. The thick line corresponds to the average of all these possible slopes, and we plot its slope ($m$) and intercept ($y_0$) and uncertainties in the top-right corner.}
    \label{fig:macs1206_1d}
\end{figure}

We produce a source plane deprojected distance map, using the ellipticity, position angle and centre from an elliptical fit to the F160W HST image of the complete multiple image of the galaxy. We forward-lens this map using {\sc lenstool}, and define 1 kpc annular apertures starting at $r=0$, measuring the average metallicity in these annuli. This approach does not include any correction for seeing, which it is known to flatten gradients \citep{Yuan2013}. We fit the data with the {\sc linmix}\footnote{https://linmix.readthedocs.io/en/latest/} package. We obtain a slope of -0.039$\pm$0.060 dex/kpc, a central metallicity 9.06$\pm$0.25 in 12+log(O/H). The data and fit are shown in Fig.~\ref{fig:macs1206_1d}.

 \subsection{Comparison with the literature}

At high-redshift, a wide range of metallicity gradients have been derived from lensing studies, which range from quite steep negative gradients \citep[e.g.][]{Jones2013,Wang2017} to positive gradients \citep[e.g.][]{Leethochawalit2016} that are usually not observed in the local Universe. However, these previous lensing studies focused on galaxies at considerably higher redshifts (1.2 $\le z \le$ 2.3) than the three objects analysed here ($z=0.6$, 0.7 and 1.0).

A better match in redshift to our sample are the \cite{Wuyts2016} and \cite{Carton2018} surveys of field galaxies. \cite{Wuyts2016} analyse a sample of 180 star-forming galaxies from the KMOS$^{3D}$ survey, from $z=0.6$ to 2.7, with stellar masses between 10$^{9.5}$ and 10$^{11.5}$ M$_\odot$ and SFR between 0.1 and 1000 M$\odot$/yr , measuring the metallicity in annuli using the N2 indicator. Most of their sample have flat gradients, with only $\approx7\%$ of the sample exhibiting positive gradients. \cite{Carton2018} analyse a sample of 84 galaxies from several MUSE GTO programmes, with stellar masses between $10^{7}$ and $10^{10.5}$ M$_\odot$ and SFR between 0.01 and 10 M$_\odot$/yr at $z=0.2-0.8$, combining several metallicity diagnostics in a 2D forward-modelling approach. They obtain a mean negative gradient of -0.039$^{+0.007}_{−0.009}$ dex/kpc, but with a larger spread in gradients than found by \cite{Wuyts2016}. AS1063-arc and A370-sys1, with redshifts of 0.6 and 0.7, are at the intersection of these two studies, and are compatible with the mean values of both. We compare M1206-sys1, at $z=1$, only with \cite{Wuyts2016}. We obtain a gradient more negative than most galaxies between $z=0.9-1.1$ (-0.006 dex/kpc) , but still compatible with \cite{Wuyts2016} within uncertainty. 

There are strong indications for the existence of a characteristic metallicity slope in low-$z$ galaxies, when the physical slope (dex/kpc) is normalised to the size of the galaxies. Both \cite{Sanchez2014} and \cite{Sanchez-Menguiano2018} find a characteristic (scaled) slope of -0.1 dex/R$_\mathrm{e}$, when the gradient is normalised to the effective radius R$_\mathrm{e}$ (see also \citealt{Ho2015} for a R$_{25}$ normalisation). At higher redshift, \citet{Carton2018}, find a steeper slope of -0.34 dex/R$_\mathrm{e}$ (for galaxies with R$_d$>3 kpc, as the ones presented here, and converting R$_\mathrm{d}$ in R$_\mathrm{e}$), although with a higher spread than found at lower redshift  ($\sigma_{\mathrm{int}}$ = 0.1 dex). 

We normalise the gradients with the values of R$_\mathrm{e}$ obtained from morphological fits (see Table~\ref{tab:sample}), obtaining $\nabla\,Z$ of -0.323$\pm$0.007, -0.636$\pm$0.011 and -0.407$\pm$0.658 dex/R$_\mathrm{e}$ for AS1063-arc, A370-sys1 and M1206-sys1, respectively. These are all significantly steeper scaled gradients than what is found for low redshift galaxies (-0.1 dex/R$_\mathrm{e}$), or for galaxies between 0.1$\le z \le$0.8 as in \cite{Carton2018}. Part of the discrepancy might be explained by errors in the R$_e$, derived using {\sc galfit}.

\subsection{Deviations from radial gradients}
\label{subsec:residuals}

Here we analyse the residuals of the metallicity maps after subtracting the fitted gradients, which we refer to as metallicity residuals. 

For AS1063-arc, when using the morphological parameters obtained with {\sc galfit}, the radial residuals are as high as 0.1 dex, when radially binned in 0.5 dex metallicity bins, but without a clear radial trend (see the bottom panel of Fig.~\ref{fig:as1063_1d}). For the fit where all variables are allowed to vary, the residuals are very low ($\leq$0.02 dex) up until 6 kpc ($\sim0.8R_e$). After this, there seems to be a trend of increasing residuals with radius. This could be caused by a flattening of the metallicity gradient at outer radii (between 0.5 to 3 R$_e$), as observed in some cases in the local Universe \citep{Sanchez-Menguiano2018}, but it would be necessary to probe the metallicity gradient further out in order to confirm this. 

As for A370-sys1, both models, with fixed or free morphological parameters, result in residuals of about $\le$0.05 dex, when the data is radially binned in bins of 0.5 dex. 

\begin{figure}
	\includegraphics[width=0.5\textwidth]{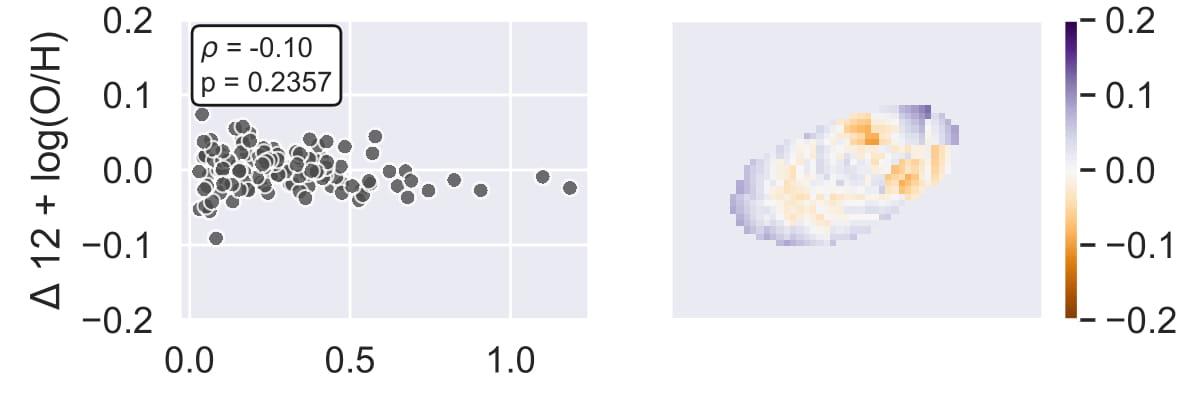}
    \includegraphics[width=0.5\textwidth]{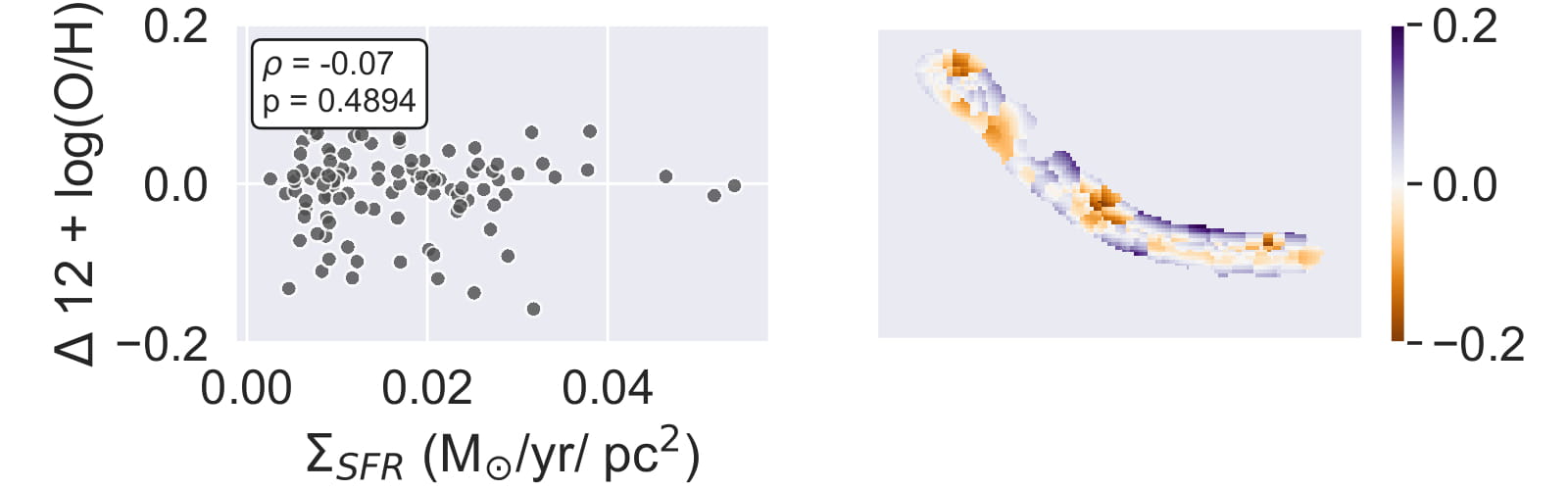}
    \caption{Metallicity gradient residuals. Top: AS1063-arc. Bottom: A370-sys1. Left: residuals after subtracting fitted gradient vs star formation rate density. The Spearman rank-order correlation coefficient and respective p value calculated for each of the two properties plotted are shown in the top-left corner of each plot. Right: 2D residuals map.}
    \label{fig:2d_residuals}
\end{figure}

In the 2D analysis of the metallicity residuals, we consider only the gradient modelled with free parameters, for simplicity. In Fig.~\ref{fig:2d_residuals} we plot the 2D residuals, as well as the residuals versus the stellar mass surface density and star-formation density. We do not see any trend with morphological features of the galaxies. We note that \cite{Erroz-Ferrer2019} in their analysis of local discs, found a metallicity increase of about $\approx0.2-0.25$ dex in \hii\, regions when compared with the surrounding metallicity. This does not appear to be the case for these z$\sim$1 galaxies, despite the fact that they do contain giant \hii\, regions, typical of high-$z$ disc galaxies.

We investigate this further by plotting the residual metallicity versus the star-formation density, also in Fig.~\ref{fig:2d_residuals}, and computing the Spearman rank correlation coefficient between these two quantities. We obtain values of $\rho=-0.1$ and -0.07, with $p$ values of 0.24 and 0.48, showing no clear correlation between the residual metallicity and the star-forming rates densities. One possible explanation for not observing the same increase in metallicity as noted in \cite{Erroz-Ferrer2019}, is the difference in spatial scales probed. Although the work presented here probes sub-kiloparsec regions, which are at $z\approx1$ only possible to study  in lensed galaxies, \cite{Erroz-Ferrer2019} observe galaxies at $<$100 pc scales, an order of magnitude smaller.

\section{Summary and Conclusions}
\label{sec:conclusions}

In this work, we made use of HST, MUSE and SINFONI data to analyse the spatially resolved properties of 3 lensed galaxies at redshifts 0.6, 0.7 and 1, at exceptionally high spatial resolution (see Fig.~\ref{fig:bin_sizes}). We derive the stellar-mass surface density using multiple HST bands. For the two lower-redshift targets, AS1063-arc and A370-sys1, we derive the gas metallicity using the line ratios (\emph{O2, O3, O32, R23}, \mbox{[O\,{\scshape iii]}}\,$\lambda$5007/4959, \Hb/\Hg) and the \cite{Maiolino2008} metallicity calibration. For M1206-sys1 only \emph{N2} was available. Using these results, we examine the resolved Star-Forming Main Sequence (rSFMS) at $z\approx1$ at sub-kiloparsec resolution, at a physical scale unattainable with un-lensed galaxies. We also explore, for the first time at $z\approx1$, the resolved Mass-Metallicity Relation (rMZR) and the resolved Fundamental Mass-Metallicity Relation (rFMZ). 

In order to fit the 2D  metallicity gradients, we develop a forward-modelling method that fits data in the image plane, correcting for seeing and lensing distortions, avoiding issues arising from combining different multiple images.

Our main results from this analysis are:

\begin{itemize}
\item We find that both the rSFMS and rMZR are in place for galaxies AS1063-arc ($z=0.6$) and A370-sys1 ($z=0.7$), although with different slopes as the ones observed in the local Universe (Fig.~\ref{fig:scaling_relations}). 
\item For these two galaxies, we also find a correlation between the residuals of the rSFMS and the rMZR ($\rho$=0.19 and 0.67, Fig.~\ref{fig:scaling_relations}), which might indicate the presence of a rFMZ. We notice however, that we find the opposite correlation (with higher rSFMS residuals corresponding to higher rMZR residuals) to what is found in other works. Moreover, the correlations are different for the two galaxies tested, which suggests that the relation evolves with redshift. A larger sample is needed in order to confirm these results.
\item We measure metallicity gradients of $-0.027\pm0.003$, $-0.019\pm0.003$ and $-0.039\pm0.060$ dex/kpc for our three targets (Table~\ref{tab:frapy_fit}). This is in agreement with what was derived for surveys at similar redshifts.
\item We find no significant deviations from an exponentially decreasing metallicity gradient (Fig.~\ref{fig:as1063_1d} and \ref{fig:a370_1d}). In particular, we find no increase or decrease of the metallicity with star-formation rate density (Fig.\ref{fig:2d_residuals}). We find a mean dispersion of the metallicity residuals of 0.01 dex for AS1063-arc and of $\approx0.05$ dex for A370-sys1.
\end{itemize}

We conclude that, although the galaxies analysed are typical high-$z$ disc galaxies, with several large \hii\, regions (clumps) and highly turbulent ionised gas, the relation between stellar mass surface density, star-formation rate surface density and metallicity at sub-kiloparsec scales observed at in local discs is already in place at $z\approx1$. Moreover, a negative metallicity gradient is already established, although with steeper scaled gradients than seen in local disc galaxies, and there are no significant metallicity deviations from a linear gradient, either due to morphological structures such as spiral arms or star-forming regions. 

The data and analysis done for this work can be found in \url{https://github.com/VeraPatricio/Resolved_Metallicity}.

\section*{Acknowledgements}

We thank Nicole Nesvadba for reducing the SINFONI data. We also thank Tiantian Yuan, Lisa Kewley and Ayan Acharyya for useful and insightful comments on how to improve this work. Finally, we thank the referee for helpful suggestions that made this work clearer.

VP is supported by the grant DFF - 4090-00079. DC acknowledges support from the ERC starting grant 336736-CALENDS. CP thanks the Alexander von Humboldt Foundation for the granting of a Bessel Research Award held at MPA. CP is also grateful to the ESO and the DFG Cluster of Excellence "Origin and Structure of the Universe" for support. JB acknowledges support by FCT/MCTES through national funds by this grant UID/FIS/04434/2019 and through the Investigador FCT Contract No. IF/01654/2014/CP1215/CT0003.

This research made use several open source {\sc python} libraries: {\sc numpy} \citep{numpy}, {\sc scipy} \citep{scipy}, {\sc matplotlib} \citep{matplotlib}, and {\sc astropy}, a community-developed core Python package for Astronomy \citep{astropy}. This research has made use of the VizieR catalogue access tool, CDS, Strasbourg, France. The original description of the VizieR service was published in A\&AS 143, 23. This work has made use of dust extinction maps from the NASA/IPAC Infrared Science Archive.



\bibliographystyle{mnras}
\bibliography{bibliography} 

\begin{thebibliography}{}
\makeatletter
\relax
\def\mn@urlcharsother{\let\do\@makeother \do\$\do\&\do\#\do\^\do\_\do\%\do\~}
\def\mn@doi{\begingroup\mn@urlcharsother \@ifnextchar [ {\mn@doi@}
  {\mn@doi@[]}}
\def\mn@doi@[#1]#2{\def\@tempa{#1}\ifx\@tempa\@empty \href
  {http://dx.doi.org/#2} {doi:#2}\else \href {http://dx.doi.org/#2} {#1}\fi
  \endgroup}
\def\mn@eprint#1#2{\mn@eprint@#1:#2::\@nil}
\def\mn@eprint@arXiv#1{\href {http://arxiv.org/abs/#1} {{\tt arXiv:#1}}}
\def\mn@eprint@dblp#1{\href {http://dblp.uni-trier.de/rec/bibtex/#1.xml}
  {dblp:#1}}
\def\mn@eprint@#1:#2:#3:#4\@nil{\def\@tempa {#1}\def\@tempb {#2}\def\@tempc
  {#3}\ifx \@tempc \@empty \let \@tempc \@tempb \let \@tempb \@tempa \fi \ifx
  \@tempb \@empty \def\@tempb {arXiv}\fi \@ifundefined
  {mn@eprint@\@tempb}{\@tempb:\@tempc}{\expandafter \expandafter \csname
  mn@eprint@\@tempb\endcsname \expandafter{\@tempc}}}

\bibitem[\protect\citeauthoryear{{Abdurro'uf} \& Akiyama}{{Abdurro'uf} \&
  Akiyama}{2018}]{Abdurrouf2018}
{Abdurro'uf} Akiyama M.,  2018, \mn@doi [\mnras] {10.1093/mnras/sty1771}, \href
  {https://ui.adsabs.harvard.edu/\#abs/2018MNRAS.479.5083A} {479, 5083}

\bibitem[\protect\citeauthoryear{{Abuter}, {Schreiber}, {Eisenhauer}, {Ott},
  {Horrobin}  \& {Gillesen}}{{Abuter} et~al.}{2006}]{Abuter2006}
{Abuter} R.,  {Schreiber} J.,  {Eisenhauer} F.,  {Ott} T.,  {Horrobin} M.,
  {Gillesen} S.,  2006, \mn@doi [New Astronomy Reviews]
  {10.1016/j.newar.2006.02.008}, \href
  {https://ui.adsabs.harvard.edu/#abs/2006NewAR..50..398A} {50, 398}

\bibitem[\protect\citeauthoryear{{Allende Prieto}, {Lambert}  \&
  {Asplund}}{{Allende Prieto} et~al.}{2001}]{AllendePrieto2001}
{Allende Prieto} C.,  {Lambert} D.~L.,   {Asplund} M.,  2001, \mn@doi [\apjl]
  {10.1086/322874}, \href {http://adsabs.harvard.edu/abs/2001ApJ...556L..63A}
  {556, L63}

\bibitem[\protect\citeauthoryear{{Astropy Collaboration} et~al.,}{{Astropy
  Collaboration} et~al.}{2013}]{astropy}
{Astropy Collaboration} et~al., 2013, \mn@doi [\aap]
  {10.1051/0004-6361/201322068}, \href
  {http://adsabs.harvard.edu/abs/2013A%26A...558A..33A} {558, A33}

\bibitem[\protect\citeauthoryear{{Bacon} et~al.,}{{Bacon}
  et~al.}{2010}]{Bacon2010}
{Bacon} R.,  et~al., 2010, in Ground-based and Airborne Instrumentation for
  Astronomy III. p. 773508, \mn@doi{10.1117/12.856027}

\bibitem[\protect\citeauthoryear{{Bacon} et~al.,}{{Bacon}
  et~al.}{2017}]{Bacon2017}
{Bacon} R.,  et~al., 2017, \mn@doi [\aap] {10.1051/0004-6361/201730833}, \href
  {http://adsabs.harvard.edu/abs/2017A%26A...608A...1B} {608, A1}

\bibitem[\protect\citeauthoryear{{Barrera-Ballesteros}
  et~al.,}{{Barrera-Ballesteros} et~al.}{2016}]{Barrera-Ballesteros2016}
{Barrera-Ballesteros} J.~K.,  et~al., 2016, \mn@doi [\mnras]
  {10.1093/mnras/stw1984}, \href
  {https://ui.adsabs.harvard.edu/\#abs/2016MNRAS.463.2513B} {463, 2513}

\bibitem[\protect\citeauthoryear{{Belfiore} et~al.,}{{Belfiore}
  et~al.}{2017}]{Belfiore2017}
{Belfiore} F.,  et~al., 2017, \mn@doi [\mnras] {10.1093/mnras/stx789}, \href
  {https://ui.adsabs.harvard.edu/\#abs/2017MNRAS.469..151B} {469, 151}

\bibitem[\protect\citeauthoryear{{Bouch{\'e}} et~al.,}{{Bouch{\'e}}
  et~al.}{2010}]{Bouche2010}
{Bouch{\'e}} N.,  et~al., 2010, \mn@doi [\apj] {10.1088/0004-637X/718/2/1001},
  \href {https://ui.adsabs.harvard.edu/\#abs/2010ApJ...718.1001B} {718, 1001}

\bibitem[\protect\citeauthoryear{{Brinchmann}, {Charlot}, {White}, {Tremonti},
  {Kauffmann}, {Heckman}  \& {Brinkmann}}{{Brinchmann}
  et~al.}{2004}]{Brinchmann2004}
{Brinchmann} J.,  {Charlot} S.,  {White} S.~D.~M.,  {Tremonti} C.,  {Kauffmann}
  G.,  {Heckman} T.,   {Brinkmann} J.,  2004, \mn@doi [\mnras]
  {10.1111/j.1365-2966.2004.07881.x}, \href
  {https://ui.adsabs.harvard.edu/\#abs/2004MNRAS.351.1151B} {351, 1151}

\bibitem[\protect\citeauthoryear{{Bruzual} \& {Charlot}}{{Bruzual} \&
  {Charlot}}{2003}]{Bruzual2003}
{Bruzual} G.,  {Charlot} S.,  2003, \mn@doi [\mnras]
  {10.1046/j.1365-8711.2003.06897.x}, \href
  {https://ui.adsabs.harvard.edu/\#abs/2003MNRAS.344.1000B} {344, 1000}

\bibitem[\protect\citeauthoryear{{Bundy} et~al.,}{{Bundy}
  et~al.}{2015}]{Bundy2015}
{Bundy} K.,  et~al., 2015, \mn@doi [\apj] {10.1088/0004-637X/798/1/7}, \href
  {http://adsabs.harvard.edu/abs/2015ApJ...798....7B} {798, 7}

\bibitem[\protect\citeauthoryear{{Calzetti}, {Armus}, {Bohlin}, {Kinney},
  {Koornneef}  \& {Storchi-Bergmann}}{{Calzetti} et~al.}{2000}]{Calzetti2000}
{Calzetti} D.,  {Armus} L.,  {Bohlin} R.~C.,  {Kinney} A.~L.,  {Koornneef} J.,
   {Storchi-Bergmann} T.,  2000, \mn@doi [\apj] {10.1086/308692}, \href
  {https://ui.adsabs.harvard.edu/#abs/2000ApJ...533..682C} {533, 682}

\bibitem[\protect\citeauthoryear{{Cappellari}}{{Cappellari}}{2017}]{Cappellari2017}
{Cappellari} M.,  2017, \mn@doi [\mnras] {10.1093/mnras/stw3020}, \href
  {http://adsabs.harvard.edu/abs/2017MNRAS.466..798C} {466, 798}

\bibitem[\protect\citeauthoryear{{Cappellari} \& {Copin}}{{Cappellari} \&
  {Copin}}{2003}]{Cappellari2003}
{Cappellari} M.,  {Copin} Y.,  2003, \mn@doi [\mnras]
  {10.1046/j.1365-8711.2003.06541.x}, \href
  {http://adsabs.harvard.edu/abs/2003MNRAS.342..345C} {342, 345}

\bibitem[\protect\citeauthoryear{{Carton} et~al.,}{{Carton}
  et~al.}{2015}]{Carton2015}
{Carton} D.,  et~al., 2015, \mn@doi [\mnras] {10.1093/mnras/stv967}, \href
  {https://ui.adsabs.harvard.edu/\#abs/2015MNRAS.451..210C} {451, 210}

\bibitem[\protect\citeauthoryear{{Carton} et~al.,}{{Carton}
  et~al.}{2017}]{Carton2017}
{Carton} D.,  et~al., 2017, \mn@doi [\mnras] {10.1093/mnras/stx545}, \href
  {https://ui.adsabs.harvard.edu/\#abs/2017MNRAS.468.2140C} {468, 2140}

\bibitem[\protect\citeauthoryear{{Carton} et~al.,}{{Carton}
  et~al.}{2018}]{Carton2018}
{Carton} D.,  et~al., 2018, \mn@doi [\mnras] {10.1093/mnras/sty1343}, \href
  {https://ui.adsabs.harvard.edu/#abs/2018MNRAS.478.4293C} {478, 4293}

\bibitem[\protect\citeauthoryear{{Chabrier}}{{Chabrier}}{2003}]{Chabrier2003}
{Chabrier} G.,  2003, \mn@doi [\pasp] {10.1086/376392}, \href
  {http://adsabs.harvard.edu/abs/2003PASP..115..763C} {115, 763}

\bibitem[\protect\citeauthoryear{{Charlot} \& {Fall}}{{Charlot} \&
  {Fall}}{2000}]{Charlot2000}
{Charlot} S.,  {Fall} S.~M.,  2000, \mn@doi [\apj] {10.1086/309250}, \href
  {http://adsabs.harvard.edu/abs/2000ApJ...539..718C} {539, 718}

\bibitem[\protect\citeauthoryear{{Conroy}, {Gunn}  \& {White}}{{Conroy}
  et~al.}{2009}]{Conroy2009}
{Conroy} C.,  {Gunn} J.~E.,   {White} M.,  2009, \mn@doi [\apj]
  {10.1088/0004-637X/699/1/486}, \href
  {http://adsabs.harvard.edu/abs/2009ApJ...699..486C} {699, 486}

\bibitem[\protect\citeauthoryear{{Croom} et~al.,}{{Croom}
  et~al.}{2012}]{Croom2012}
{Croom} S.~M.,  et~al., 2012, \mn@doi [\mnras]
  {10.1111/j.1365-2966.2011.20365.x}, \href
  {https://ui.adsabs.harvard.edu/\#abs/2012MNRAS.421..872C} {421, 872}

\bibitem[\protect\citeauthoryear{{D'Eugenio}, {Colless}, {Groves}, {Bian}  \&
  {Barone}}{{D'Eugenio} et~al.}{2018}]{DEugenio2018}
{D'Eugenio} F.,  {Colless} M.,  {Groves} B.,  {Bian} F.,   {Barone} T.~M.,
  2018, \mn@doi [\mnras] {10.1093/mnras/sty1424}, \href
  {http://adsabs.harvard.edu/abs/2018MNRAS.479.1807D} {479, 1807}

\bibitem[\protect\citeauthoryear{{Dav{\'e}}, {Finlator}  \&
  {Oppenheimer}}{{Dav{\'e}} et~al.}{2012}]{Dave2012}
{Dav{\'e}} R.,  {Finlator} K.,   {Oppenheimer} B.~D.,  2012, \mn@doi [\mnras]
  {10.1111/j.1365-2966.2011.20148.x}, \href
  {https://ui.adsabs.harvard.edu/\#abs/2012MNRAS.421...98D} {421, 98}

\bibitem[\protect\citeauthoryear{{Davies}}{{Davies}}{2007}]{Davies2007}
{Davies} R.~I.,  2007, \mn@doi [\mnras] {10.1111/j.1365-2966.2006.11383.x},
  \href {https://ui.adsabs.harvard.edu/#abs/2007MNRAS.375.1099D} {375, 1099}

\bibitem[\protect\citeauthoryear{{Dutton}, {van den Bosch}  \&
  {Dekel}}{{Dutton} et~al.}{2010}]{Dutton2010}
{Dutton} A.~A.,  {van den Bosch} F.~C.,   {Dekel} A.,  2010, \mn@doi [\mnras]
  {10.1111/j.1365-2966.2010.16620.x}, \href
  {https://ui.adsabs.harvard.edu/\#abs/2010MNRAS.405.1690D} {405, 1690}

\bibitem[\protect\citeauthoryear{{Eisenhauer} et~al.,}{{Eisenhauer}
  et~al.}{2003}]{Eisenhauer2003}
{Eisenhauer} F.,  et~al., 2003, in {Iye} M.,  {Moorwood} A.~F.~M.,  eds,
  \procspie Vol. 4841, Instrument Design and Performance for Optical/Infrared
  Ground-based Telescopes. pp 1548--1561 (\mn@eprint {} {astro-ph/0306191}),
  \mn@doi{10.1117/12.459468}

\bibitem[\protect\citeauthoryear{{Erb}, {Shapley}, {Pettini}, {Steidel},
  {Reddy}  \& {Adelberger}}{{Erb} et~al.}{2006}]{Erb2006}
{Erb} D.~K.,  {Shapley} A.~E.,  {Pettini} M.,  {Steidel} C.~C.,  {Reddy} N.~A.,
    {Adelberger} K.~L.,  2006, \mn@doi [\apj] {10.1086/503623}, \href
  {https://ui.adsabs.harvard.edu/abs/2006ApJ...644..813E} {644, 813}

\bibitem[\protect\citeauthoryear{{Erroz-Ferrer} et~al.,}{{Erroz-Ferrer}
  et~al.}{2019}]{Erroz-Ferrer2019}
{Erroz-Ferrer} S.,  et~al., 2019, \mn@doi [\mnras] {10.1093/mnras/stz194},
  \href {https://ui.adsabs.harvard.edu/\#abs/2019MNRAS.tmp..200E} {p.~200}

\bibitem[\protect\citeauthoryear{{Foreman-Mackey}, {Hogg}, {Lang}  \&
  {Goodman}}{{Foreman-Mackey} et~al.}{2013}]{Foreman-Mackey2013}
{Foreman-Mackey} D.,  {Hogg} D.~W.,  {Lang} D.,   {Goodman} J.,  2013, \mn@doi
  [\pasp] {10.1086/670067}, \href
  {http://adsabs.harvard.edu/abs/2013PASP..125..306F} {125, 306}

\bibitem[\protect\citeauthoryear{{F{\"o}rster Schreiber} et~al.,}{{F{\"o}rster
  Schreiber} et~al.}{2009}]{Forster-Schreiber2009}
{F{\"o}rster Schreiber} N.~M.,  et~al., 2009, \mn@doi [\apj]
  {10.1088/0004-637X/706/2/1364}, \href
  {https://ui.adsabs.harvard.edu/#abs/2009ApJ...706.1364F} {706, 1364}

\bibitem[\protect\citeauthoryear{{Ho}, {Kudritzki}, {Kewley}, {Zahid},
  {Dopita}, {Bresolin}  \& {Rupke}}{{Ho} et~al.}{2015}]{Ho2015}
{Ho} I.~T.,  {Kudritzki} R.-P.,  {Kewley} L.~J.,  {Zahid} H.~J.,  {Dopita}
  M.~A.,  {Bresolin} F.,   {Rupke} D. S.~N.,  2015, \mn@doi [\mnras]
  {10.1093/mnras/stv067}, \href
  {https://ui.adsabs.harvard.edu/#abs/2015MNRAS.448.2030H} {448, 2030}

\bibitem[\protect\citeauthoryear{Hunter}{Hunter}{2007}]{matplotlib}
Hunter J.~D.,  2007, \mn@doi [Computing in Science Engineering]
  {10.1109/MCSE.2007.55}, 9, 90

\bibitem[\protect\citeauthoryear{{Jones}, {Ellis}, {Richard}  \&
  {Jullo}}{{Jones} et~al.}{2013}]{Jones2013}
{Jones} T.,  {Ellis} R.~S.,  {Richard} J.,   {Jullo} E.,  2013, \mn@doi [\apj]
  {10.1088/0004-637X/765/1/48}, \href
  {https://ui.adsabs.harvard.edu/#abs/2013ApJ...765...48J} {765, 48}

\bibitem[\protect\citeauthoryear{Jones, Oliphant, Peterson  et~al.}{Jones
  et~al.}{01  }]{scipy}
Jones E.,  Oliphant T.,  Peterson P.,   et~al., 2001--, {SciPy}: Open source
  scientific tools for {Python}, \url {http://www.scipy.org/}

\bibitem[\protect\citeauthoryear{{Jullo}, {Kneib}, {Limousin},
  {El{\'{\i}}asd{\'o}ttir}, {Marshall}  \& {Verdugo}}{{Jullo}
  et~al.}{2007}]{Jullo2007}
{Jullo} E.,  {Kneib} J.-P.,  {Limousin} M.,  {El{\'{\i}}asd{\'o}ttir} {\'A}.,
  {Marshall} P.~J.,   {Verdugo} T.,  2007, \mn@doi [New Journal of Physics]
  {10.1088/1367-2630/9/12/447}, \href
  {http://cdsads.u-strasbg.fr/abs/2007NJPh....9..447J} {9, 447}

\bibitem[\protect\citeauthoryear{{Karman} et~al.,}{{Karman}
  et~al.}{2015}]{Karman2015}
{Karman} W.,  et~al., 2015, \mn@doi [\aap] {10.1051/0004-6361/201424962}, \href
  {http://adsabs.harvard.edu/abs/2015A%26A...574A..11K} {574, A11}

\bibitem[\protect\citeauthoryear{{Kelly}}{{Kelly}}{2007}]{Kelly2007}
{Kelly} B.~C.,  2007, \mn@doi [\apj] {10.1086/519947}, \href
  {http://adsabs.harvard.edu/abs/2007ApJ...665.1489K} {665, 1489}

\bibitem[\protect\citeauthoryear{{Kennicutt}}{{Kennicutt}}{1998}]{Kennicutt1998}
{Kennicutt} Jr. R.~C.,  1998, \mn@doi [\araa] {10.1146/annurev.astro.36.1.189},
  \href {http://adsabs.harvard.edu/abs/1998ARA%26A..36..189K} {36, 189}

\bibitem[\protect\citeauthoryear{{Kewley} \& {Dopita}}{{Kewley} \&
  {Dopita}}{2002}]{Kewley2002}
{Kewley} L.~J.,  {Dopita} M.~A.,  2002, \mn@doi [\apjs] {10.1086/341326}, \href
  {http://adsabs.harvard.edu/abs/2002ApJS..142...35K} {142, 35}

\bibitem[\protect\citeauthoryear{{Kriek}, {van Dokkum}, {Labb{\'e}}, {Franx},
  {Illingworth}, {Marchesini}  \& {Quadri}}{{Kriek} et~al.}{2009}]{Kriek2009}
{Kriek} M.,  {van Dokkum} P.~G.,  {Labb{\'e}} I.,  {Franx} M.,  {Illingworth}
  G.~D.,  {Marchesini} D.,   {Quadri} R.~F.,  2009, \mn@doi [\apj]
  {10.1088/0004-637X/700/1/221}, \href
  {http://adsabs.harvard.edu/abs/2009ApJ...700..221K} {700, 221}

\bibitem[\protect\citeauthoryear{{Lara-L{\'o}pez} et~al.,}{{Lara-L{\'o}pez}
  et~al.}{2010}]{Lara2010}
{Lara-L{\'o}pez} M.~A.,  et~al., 2010, \mn@doi [\aap]
  {10.1051/0004-6361/201014803}, \href
  {http://adsabs.harvard.edu/abs/2010A%26A...521L..53L} {521, L53}

\bibitem[\protect\citeauthoryear{{Larson}}{{Larson}}{1976}]{Larson1976}
{Larson} R.~B.,  1976, \mn@doi [\mnras] {10.1093/mnras/176.1.31}, \href
  {https://ui.adsabs.harvard.edu/\#abs/1976MNRAS.176...31L} {176, 31}

\bibitem[\protect\citeauthoryear{{Leethochawalit}, {Jones}, {Ellis}, {Stark},
  {Richard}, {Zitrin}  \& {Auger}}{{Leethochawalit}
  et~al.}{2016}]{Leethochawalit2016}
{Leethochawalit} N.,  {Jones} T.~A.,  {Ellis} R.~S.,  {Stark} D.~P.,  {Richard}
  J.,  {Zitrin} A.,   {Auger} M.,  2016, \mn@doi [\apj]
  {10.3847/0004-637X/820/2/84}, \href
  {https://ui.adsabs.harvard.edu/#abs/2016ApJ...820...84L} {820, 84}

\bibitem[\protect\citeauthoryear{{Lilly}, {Peng}, {Renzini}  \&
  {Carollo}}{{Lilly} et~al.}{2013}]{Lilly2013}
{Lilly} S.~J.,  {Peng} Y.,  {Renzini} A.,   {Carollo} C.~M.,  2013, in {Sun}
  W.~H.,  {Xu} C.~K.,  {Scoville} N.~Z.,   {Sanders} D.~B.,  eds,  Astronomical
  Society of the Pacific Conference Series Vol. 477, Galaxy Mergers in an
  Evolving Universe. p.~11

\bibitem[\protect\citeauthoryear{{Maiolino}, {Nagao}, {Grazian}, {Cocchia},
  {Marconi}, {Mannucci}, {Cimatti}  \& {Pipino}}{{Maiolino}
  et~al.}{2008}]{Maiolino2008}
{Maiolino} R.,  {Nagao} T.,  {Grazian} A.,  {Cocchia} F.,  {Marconi} A.,
  {Mannucci} F.,  {Cimatti} A.,   {Pipino} A.,  2008, \mn@doi [\aap]
  {10.1051/0004-6361:200809678}, \href
  {http://adsabs.harvard.edu/abs/2008A%26A...488..463M} {488, 463}

\bibitem[\protect\citeauthoryear{{Mannucci}, {Cresci}, {Maiolino}, {Marconi}
  \& {Gnerucci}}{{Mannucci} et~al.}{2010}]{Mannucci2010}
{Mannucci} F.,  {Cresci} G.,  {Maiolino} R.,  {Marconi} A.,   {Gnerucci} A.,
  2010, \mn@doi [\mnras] {10.1111/j.1365-2966.2010.17291.x}, \href
  {http://adsabs.harvard.edu/abs/2010MNRAS.408.2115M} {408, 2115}

\bibitem[\protect\citeauthoryear{{Marino} et~al.,}{{Marino}
  et~al.}{2013}]{Marino2013}
{Marino} R.~A.,  et~al., 2013, \mn@doi [\aap] {10.1051/0004-6361/201321956},
  \href {https://ui.adsabs.harvard.edu/abs/2013A&A...559A.114M} {559, A114}

\bibitem[\protect\citeauthoryear{{Mott}, {Spitoni}  \& {Matteucci}}{{Mott}
  et~al.}{2013}]{Mott2013}
{Mott} A.,  {Spitoni} E.,   {Matteucci} F.,  2013, \mn@doi [\mnras]
  {10.1093/mnras/stt1495}, \href
  {https://ui.adsabs.harvard.edu/\#abs/2013MNRAS.435.2918M} {435, 2918}

\bibitem[\protect\citeauthoryear{{Paalvast} et~al.,}{{Paalvast}
  et~al.}{2018}]{Paalvast2018}
{Paalvast} M.,  et~al., 2018, \mn@doi [\aap] {10.1051/0004-6361/201832866},
  \href {https://ui.adsabs.harvard.edu/abs/2018A&A...618A..40P} {618, A40}

\bibitem[\protect\citeauthoryear{{Patr{\'\i}cio} et~al.,}{{Patr{\'\i}cio}
  et~al.}{2018}]{Patricio2018}
{Patr{\'\i}cio} V.,  et~al., 2018, \mn@doi [\mnras] {10.1093/mnras/sty555},
  \href {https://ui.adsabs.harvard.edu/abs/2018MNRAS.477...18P} {477, 18}

\bibitem[\protect\citeauthoryear{{Peng}, {Ho}, {Impey}  \& {Rix}}{{Peng}
  et~al.}{2010}]{Peng2010}
{Peng} C.~Y.,  {Ho} L.~C.,  {Impey} C.~D.,   {Rix} H.-W.,  2010, \mn@doi [\aj]
  {10.1088/0004-6256/139/6/2097}, \href
  {http://adsabs.harvard.edu/abs/2010AJ....139.2097P} {139, 2097}

\bibitem[\protect\citeauthoryear{{Pettini} \& {Pagel}}{{Pettini} \&
  {Pagel}}{2004}]{Pettini2004}
{Pettini} M.,  {Pagel} B. E.~J.,  2004, \mn@doi [\mnras]
  {10.1111/j.1365-2966.2004.07591.x}, \href
  {https://ui.adsabs.harvard.edu/abs/2004MNRAS.348L..59P} {348, L59}

\bibitem[\protect\citeauthoryear{{Pilkington} et~al.,}{{Pilkington}
  et~al.}{2012}]{Pilkington2012}
{Pilkington} K.,  et~al., 2012, \mn@doi [\aap] {10.1051/0004-6361/201117466},
  \href {https://ui.adsabs.harvard.edu/\#abs/2012A&A...540A..56P} {540, A56}

\bibitem[\protect\citeauthoryear{{Pilyugin}, {Grebel}  \&
  {Zinchenko}}{{Pilyugin} et~al.}{2015}]{Pilyugin2015}
{Pilyugin} L.~S.,  {Grebel} E.~K.,   {Zinchenko} I.~A.,  2015, \mn@doi [\mnras]
  {10.1093/mnras/stv932}, \href
  {https://ui.adsabs.harvard.edu/\#abs/2015MNRAS.450.3254P} {450, 3254}

\bibitem[\protect\citeauthoryear{{Rosales-Ortega}, {S{\'a}nchez},
  {Iglesias-P{\'a}ramo}, {D{\'\i}az}, {V{\'\i}lchez}, {Bland-Hawthorn},
  {Husemann}  \& {Mast}}{{Rosales-Ortega} et~al.}{2012}]{Rosales-Ortega2012}
{Rosales-Ortega} F.~F.,  {S{\'a}nchez} S.~F.,  {Iglesias-P{\'a}ramo} J.,
  {D{\'\i}az} A.~I.,  {V{\'\i}lchez} J.~M.,  {Bland-Hawthorn} J.,  {Husemann}
  B.,   {Mast} D.,  2012, \mn@doi [\apj] {10.1088/2041-8205/756/2/L31}, \href
  {https://ui.adsabs.harvard.edu/\#abs/2012ApJ...756L..31R} {756, L31}

\bibitem[\protect\citeauthoryear{{S{\'a}nchez-Menguiano}
  et~al.,}{{S{\'a}nchez-Menguiano} et~al.}{2016}]{Sanchez-Menguiano2016}
{S{\'a}nchez-Menguiano} L.,  et~al., 2016, \mn@doi [\aap]
  {10.1051/0004-6361/201527450}, \href
  {https://ui.adsabs.harvard.edu/\#abs/2016A&A...587A..70S} {587, A70}

\bibitem[\protect\citeauthoryear{{S{\'a}nchez-Menguiano}
  et~al.,}{{S{\'a}nchez-Menguiano} et~al.}{2018}]{Sanchez-Menguiano2018}
{S{\'a}nchez-Menguiano} L.,  et~al., 2018, \mn@doi [\aap]
  {10.1051/0004-6361/201731486}, \href
  {https://ui.adsabs.harvard.edu/#abs/2018A&A...609A.119S} {609, A119}

\bibitem[\protect\citeauthoryear{{S{\'a}nchez} et~al.,}{{S{\'a}nchez}
  et~al.}{2013}]{Sanchez2013}
{S{\'a}nchez} S.~F.,  et~al., 2013, \mn@doi [\aap]
  {10.1051/0004-6361/201220669}, \href
  {https://ui.adsabs.harvard.edu/\#abs/2013A&A...554A..58S} {554, A58}

\bibitem[\protect\citeauthoryear{{S{\'a}nchez} et~al.,}{{S{\'a}nchez}
  et~al.}{2014}]{Sanchez2014}
{S{\'a}nchez} S.~F.,  et~al., 2014, \mn@doi [\aap]
  {10.1051/0004-6361/201322343}, \href
  {https://ui.adsabs.harvard.edu/#abs/2014A&A...563A..49S} {563, A49}

\bibitem[\protect\citeauthoryear{{S{\'a}nchez} et~al.,}{{S{\'a}nchez}
  et~al.}{2017}]{Sanchez2017}
{S{\'a}nchez} S.~F.,  et~al., 2017, \mn@doi [\mnras] {10.1093/mnras/stx808},
  \href {https://ui.adsabs.harvard.edu/abs/2017MNRAS.469.2121S} {469, 2121}

\bibitem[\protect\citeauthoryear{{Schaye} et~al.,}{{Schaye}
  et~al.}{2010}]{Schaye2010}
{Schaye} J.,  et~al., 2010, \mn@doi [\mnras]
  {10.1111/j.1365-2966.2009.16029.x}, \href
  {https://ui.adsabs.harvard.edu/\#abs/2010MNRAS.402.1536S} {402, 1536}

\bibitem[\protect\citeauthoryear{{Schlafly} \& {Finkbeiner}}{{Schlafly} \&
  {Finkbeiner}}{2011}]{Schlafly2011}
{Schlafly} E.~F.,  {Finkbeiner} D.~P.,  2011, \mn@doi [\apj]
  {10.1088/0004-637X/737/2/103}, \href
  {https://ui.adsabs.harvard.edu/abs/2011ApJ...737..103S} {737, 103}

\bibitem[\protect\citeauthoryear{{Sharma}, {Richard}, {Yuan}, {Gupta},
  {Kewley}, {Patr{\'{\i}}cio}, {Leethochawalit}  \& {Jones}}{{Sharma}
  et~al.}{2018}]{Sharma2018}
{Sharma} S.,  {Richard} J.,  {Yuan} T.,  {Gupta} A.,  {Kewley} L.,
  {Patr{\'{\i}}cio} V.,  {Leethochawalit} N.,   {Jones} T.~A.,  2018, \mn@doi
  [\mnras] {10.1093/mnras/sty2352}, \href
  {http://adsabs.harvard.edu/abs/2018MNRAS.481.1427S} {481, 1427}

\bibitem[\protect\citeauthoryear{{Shirazi}, {Brinchmann}  \&
  {Rahmati}}{{Shirazi} et~al.}{2014}]{Shirazi2014}
{Shirazi} M.,  {Brinchmann} J.,   {Rahmati} A.,  2014, \mn@doi [\apj]
  {10.1088/0004-637X/787/2/120}, \href
  {https://ui.adsabs.harvard.edu/abs/2014ApJ...787..120S} {787, 120}

\bibitem[\protect\citeauthoryear{{Soto}, {Lilly}, {Bacon}, {Richard}  \&
  {Conseil}}{{Soto} et~al.}{2016}]{Soto2016}
{Soto} K.~T.,  {Lilly} S.~J.,  {Bacon} R.,  {Richard} J.,   {Conseil} S.,
  2016, \mn@doi [\mnras] {10.1093/mnras/stw474}, \href
  {http://adsabs.harvard.edu/abs/2016MNRAS.458.3210S} {458, 3210}

\bibitem[\protect\citeauthoryear{{Speagle}, {Steinhardt}, {Capak}  \&
  {Silverman}}{{Speagle} et~al.}{2014}]{Speagle2014}
{Speagle} J.~S.,  {Steinhardt} C.~L.,  {Capak} P.~L.,   {Silverman} J.~D.,
  2014, \mn@doi [The Astrophysical Journal Supplement Series]
  {10.1088/0067-0049/214/2/15}, \href
  {https://ui.adsabs.harvard.edu/\#abs/2014ApJS..214...15S} {214, 15}

\bibitem[\protect\citeauthoryear{{Storey} \& {Hummer}}{{Storey} \&
  {Hummer}}{1995}]{Storey1995}
{Storey} P.~J.,  {Hummer} D.~G.,  1995, \mn@doi [\mnras]
  {10.1093/mnras/272.1.41}, \href
  {http://adsabs.harvard.edu/abs/1995MNRAS.272...41S} {272, 41}

\bibitem[\protect\citeauthoryear{{Trayford} \& {Schaye}}{{Trayford} \&
  {Schaye}}{2018}]{Trayford2018}
{Trayford} J.~W.,  {Schaye} J.,  2018, arXiv e-prints, \href
  {https://ui.adsabs.harvard.edu/\#abs/2018arXiv181206984T} {p.
  arXiv:1812.06984}

\bibitem[\protect\citeauthoryear{{Tremonti} et~al.,}{{Tremonti}
  et~al.}{2004}]{Tremonti2004}
{Tremonti} C.~A.,  et~al., 2004, \mn@doi [\apj] {10.1086/423264}, \href
  {http://adsabs.harvard.edu/abs/2004ApJ...613..898T} {613, 898}

\bibitem[\protect\citeauthoryear{{Valdes}, {Gupta}, {Rose}, {Singh}  \&
  {Bell}}{{Valdes} et~al.}{2004}]{Valdes2004}
{Valdes} F.,  {Gupta} R.,  {Rose} J.~A.,  {Singh} H.~P.,   {Bell} D.~J.,  2004,
  \mn@doi [\apjs] {10.1086/386343}, \href
  {http://adsabs.harvard.edu/abs/2004ApJS..152..251V} {152, 251}

\bibitem[\protect\citeauthoryear{{Wang} et~al.,}{{Wang}
  et~al.}{2017}]{Wang2017}
{Wang} X.,  et~al., 2017, \mn@doi [\apj] {10.3847/1538-4357/aa603c}, \href
  {https://ui.adsabs.harvard.edu/\#abs/2017ApJ...837...89W} {837, 89}

\bibitem[\protect\citeauthoryear{{Weilbacher}, {Streicher}  \&
  {Palsa}}{{Weilbacher} et~al.}{2016}]{Weilbacher2016}
{Weilbacher} P.~M.,  {Streicher} O.,   {Palsa} R.,  2016, {MUSE-DRP: MUSE Data
  Reduction Pipeline}, Astrophysics Source Code Library (\mn@eprint {ascl}
  {1610.004})

\bibitem[\protect\citeauthoryear{{Wesson}}{{Wesson}}{2016}]{Wesson2016}
{Wesson} R.,  2016, \mn@doi [\mnras] {10.1093/mnras/stv2946}, \href
  {http://adsabs.harvard.edu/abs/2016MNRAS.456.3774W} {456, 3774}

\bibitem[\protect\citeauthoryear{{Whitaker}, {van Dokkum}, {Brammer}  \&
  {Franx}}{{Whitaker} et~al.}{2012}]{Whitaker2012}
{Whitaker} K.~E.,  {van Dokkum} P.~G.,  {Brammer} G.,   {Franx} M.,  2012,
  \mn@doi [\apj] {10.1088/2041-8205/754/2/L29}, \href
  {https://ui.adsabs.harvard.edu/\#abs/2012ApJ...754L..29W} {754, L29}

\bibitem[\protect\citeauthoryear{{Wuyts} et~al.,}{{Wuyts}
  et~al.}{2013}]{Wuyts2013}
{Wuyts} S.,  et~al., 2013, \mn@doi [\apj] {10.1088/0004-637X/779/2/135}, \href
  {https://ui.adsabs.harvard.edu/\#abs/2013ApJ...779..135W} {779, 135}

\bibitem[\protect\citeauthoryear{{Wuyts} et~al.,}{{Wuyts}
  et~al.}{2016}]{Wuyts2016}
{Wuyts} E.,  et~al., 2016, \mn@doi [\apj] {10.3847/0004-637X/827/1/74}, \href
  {https://ui.adsabs.harvard.edu/#abs/2016ApJ...827...74W} {827, 74}

\bibitem[\protect\citeauthoryear{{Yuan}, {Kewley}  \& {Rich}}{{Yuan}
  et~al.}{2013}]{Yuan2013}
{Yuan} T.-T.,  {Kewley} L.~J.,   {Rich} J.,  2013, \mn@doi [\apj]
  {10.1088/0004-637X/767/2/106}, \href
  {http://adsabs.harvard.edu/abs/2013ApJ...767..106Y} {767, 106}

\bibitem[\protect\citeauthoryear{van~der Walt, Colbert  \& Varoquaux}{van~der
  Walt et~al.}{2011}]{numpy}
van~der Walt S.,  Colbert S.~C.,   Varoquaux G.,  2011, \mn@doi [Computing in
  Science Engineering] {10.1109/MCSE.2011.37}, 13, 22

\makeatother
\end{thebibliography}



\appendix

\section{Emission line and line ratios maps}
\label{app:line_maps}

\begin{figure*}
    \includegraphics[width=\textwidth]{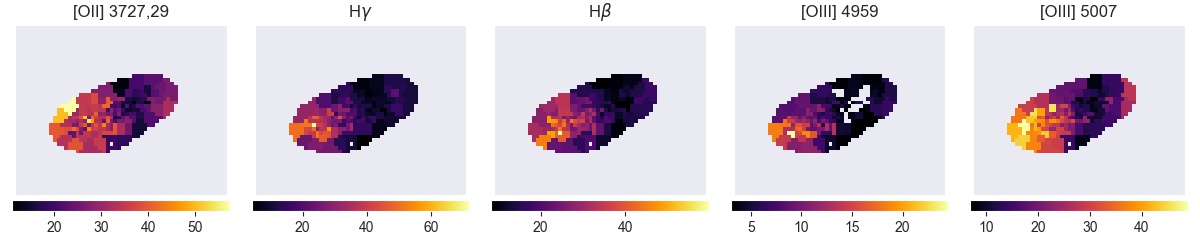}
    \includegraphics[width=\textwidth]{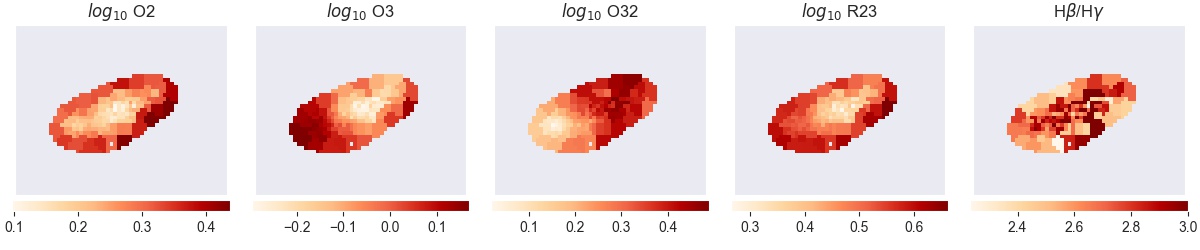}
    \includegraphics[width=\textwidth]{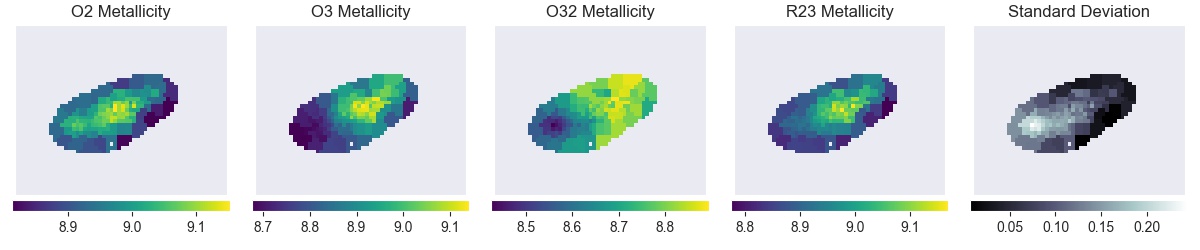}
    \caption{AS1063-arc. Top panels: Signal-to-noise maps of the emission lines used in this work. Middle row: line ratio maps (in logarithmic scale), without dust attenuation correction. Bottom row: Metallicity maps derived using the \citet{Maiolino2008} calibrations and each diagnostic individually. On the bottom-right panel, we plot the standard deviation of these values for each bin.}
    \label{fig:as1063_emission_lines}
\end{figure*}

\begin{figure*}
    \includegraphics[width=\textwidth]{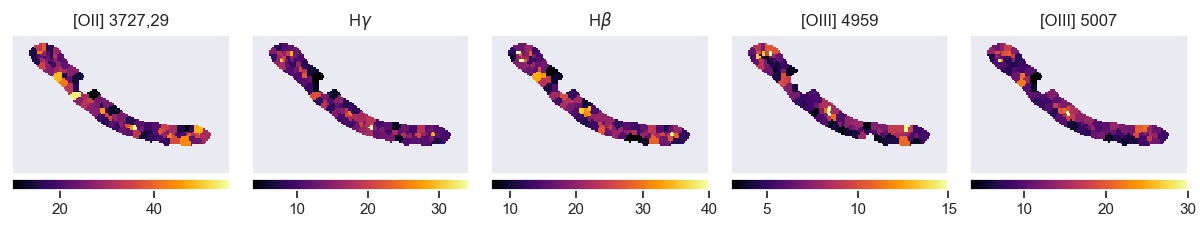}
    \includegraphics[width=\textwidth]{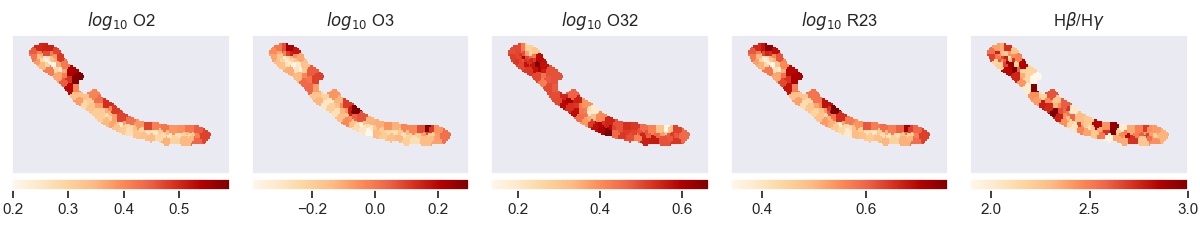}
    \includegraphics[width=\textwidth]{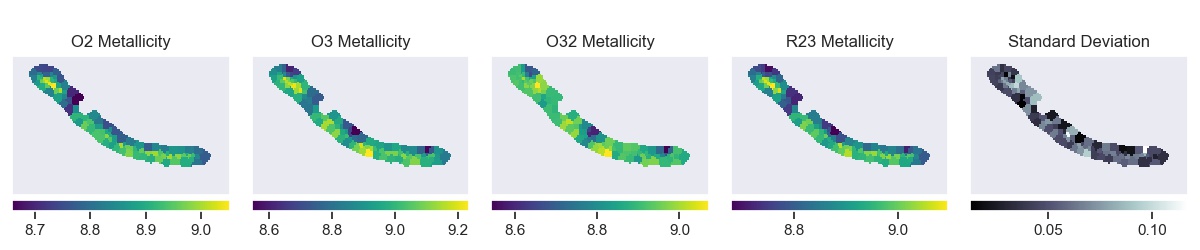}
    \caption{As Fig.~\ref{fig:as1063_emission_lines} but for A370-sys1.}
    \label{fig:a370_emission_lines}
\end{figure*}

We present the signal-to-noise ratio maps of the emission lines used to derive metallicity  (\oii, \Hb, \oiiia, \oiiib, \nii\, and \Ha) in the top panels of Fig.~\ref{fig:as1063_emission_lines} and \ref{fig:a370_emission_lines}. For \oii, we plot the sum of the doublet. The signal to noise ratio was calculated using the flux and uncertainties measured with {\sc alfa}, as detailed in Section~\ref{sec:metallicity}.

Using these maps, without including any dust correction, we calculate the individual line ratios used in this work (middle rows of Fig.~\ref{fig:as1063_emission_lines} and \ref{fig:a370_emission_lines}). Using these and the \cite{Maiolino2008} calibrations, we calculate the metallicity maps for each individual diagnostic. We notice that we obtain the largest discrepancies with \emph{O32}, an ionisation sensitive diagnostic. We also measure the dispersion in metallicity for each bin, calculating the standard deviation in each bin between of all metallicity maps.

\begin{figure}
    \includegraphics[width=0.5\textwidth]{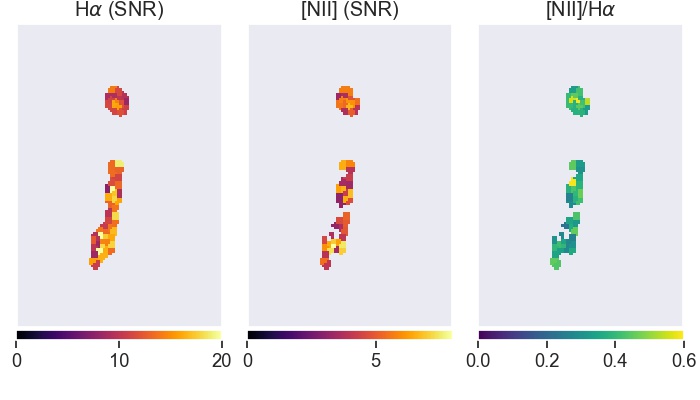}
    \caption{M1206-sys1 signal-to-noise ratios maps of \Ha\, and \nii\, (left and middle panels) and ratio of the two (right panel).}
    \label{fig:macs1206_emission_lines}
\end{figure}

For M1206-sys1, since we have only one line ratio available, we present only the signal-to-noise ratio maps of the two lines used (\Ha\, and \nii) and the ratio of the two in Fig.~\ref{fig:macs1206_emission_lines}.

\section{Comparison of metallicity derived from different line sets}
\label{app:integrated_metallicity}

\begin{table*}
\caption{Comparison between metallicities derived in P18, using the full set of lines available in MUSE and the \citet{Charlot2000} extinction law, and the metallicities derived using only the strongest lines and the \citet{Calzetti2000} extinction law. Z: the metallicity, in \met; E(B-V): dust attenuation in magnitudes; $\tau_v$: dust attenuation (adimensional).}
\label{tab:new_met}
\centering
\tabcolsep=0.2cm
\begin{tabular}{|lccccc|} 
\hline
Object        & Line Ratios & \multicolumn{2}{c}{\citet{Calzetti2000}} & \multicolumn{2}{c}{\citet{Charlot2000}} \\
         	  &             & Z &   E(B-V) 							& Z &  $\tau_v$  \\
\hline\hline
AS1063-arc & P18 & - & - & 8.82$\pm$0.02 & 1.09$\pm$0.12  \\
AS1063-arc & \emph{O2, O3, O32, R23}, \Hb/\Hg & 8.76$\pm$0.10 & 0.46$\pm$0.09 & 8.75$\pm$0.10 & 1.11$\pm$0.20 \\
\hline
A370-sys1 & P18 & - & - & 8.88$\pm$0.02 & 0.44$\pm$0.11 \\
A370-sys1 & \emph{O2, O3, O32, R23}, \Hb/\Hg  & 8.81$\pm$0.17  &   0.38$\pm$0.19 & 8.80$\pm$0.16 & 0.88$\pm$0.48 \\
\hline
M1206-sys1 & P18 & - & - & 8.91$\pm$0.06 & 0.74$\pm$0.33 \\
M1206-sys1 & \emph{O2, Ne3O2, N2}, \Hg/H7, \Hg/\Hd\,   & 8.89$\pm$0.05 & 0.92$\pm$0.11 & 8.91$\pm$0.05 & 1.86$\pm$0.21 \\
M1206-sys1 & \emph{O2, Ne3O2, N2}, \Hg/H7, \Hg/\Hd\,, \Ha/\Hd\,, \Ha/\Hg  & 8.87$\pm$0.07 & 1.00$\pm$+0.00 & 8.88$\pm$0.08 & 2.00$\pm$0.00 \\
M1206-sys1 & \emph{N2}         & 8.94$\pm$0.07 & - & 8.94$\pm$0.07 & - \\
\hline
\end{tabular}\\
\end{table*}

We compare the metallicity derived in this work using only the strongest lines, with the one obtained in \cite{Patricio2018} (hereafter P18) from integrated spectra, where faint lines were also included (\neiii,\,\Hg,\,\Hd,\, and H7). For M1206-sys1, we also test the consistency of the results derived using MUSE and SINFONI data or just SINFONI data.

Besides all the metallicity dependent rations presented in Section~\ref{subsec:method}, we also included here the \emph{Ne3O2} (\neiii/\oii) ratio and the following metallicity independent ratios:

\begin{description}[align=left,labelwidth=1cm]
\item [\Ha/\Hg]  6.113
\item [\Ha/\Hd]  11.057
\item [\Ha/H7] 18.004
\item [\Hb/\Hg] 2.135
\item [\Hg/H7] 6.288
\item [\Hg/Hd] 1.809
\item [\Hd/H7] 1.628
\item [\mbox{[O\,{\scshape III]}}\,$\lambda$5007/4959] 2.98
\end{description}

In P18, 10 line ratios were used to derive the integrated metallicity of AS1063-arc and A370-sys1 (see Table~\ref{tab:new_met}). In this work, only 5 ratios (\emph{O2, O3, O32, R23}, and \Hb/\Hg) are available to study the resolved metallicity and we re-derived the integrated metallicity using only those 5 ratios and compare it with the previous values. The new metallicity and extinction are presented in Table~\ref{tab:new_met}. 

We obtain slightly lower metallicities for AS1073-arc -- from 8.82$\pm$0.02 in P18 using 10 line ratios, to 8.75$\pm$0.10 in \met\, -- and A370-sys1 -- \met\,= 8.88$\pm$0.02 in P18 and 8.83$\pm$0.15 in this work --  but that are compatible within uncertainty. Indeed, the uncertainty of the metallicities derived in this work are considerably higher (and more realistic) than in P18, reflecting both the use of less constraints and the addition of the continuum subtraction uncertainty to the line flux errors. A similar trend is seen with the values of $\tau_v$, the extinction factor obtained with the \citet{Charlot2000} law, that are higher than in P18. As previously described, the chosen extinction law has a very small impact in the metallicity derived, about 0.01 dex, much smaller than the associated uncertainties. It seems then possible to obtain metallicities comparable as the ones derived using a larger set of line ratios, using only the 6 line ratios involving the strongest lines, although with a higher associated uncertainty.

For M1206-sys1, the MUSE data only covers \oii\, and \neiii, as well as several weak Balmer lines, \Hg, \Hd\, and H7. However, using SINFONI, both \Ha\, and \nii\, can be utilised. We first start to test whether the \emph{N2} diagnostic gives compatible results with the ones presented in P18, using \emph{Ne3O2} and \emph{O2} (see Table~\ref{tab:new_met}). We obtain a metallicity of 8.89$\pm$0.8 in \met, compatible with what was previously derived not including \emph{N2}. However, the $\tau_v$ obtained is quite higher, indicating some possible remaining issues with the flux calibration between MUSE and SINFONI data (we remind the reader that the method used here to determine metallicity uses all lines to determine extinction). Indeed, if we add more line ratios involving \Ha\, and other Balmer lines in the MUSE data, the dust attenuation values obtained are clustered around our highest allowed extinction, much higher than what is obtained with only the MUSE data, and surprisingly high (E(V-B)$>1$ mag). We conclude that our flux calibration between the two data sets is not accurate enough to allow to robustly determine the extinction combining \Ha\, with other Balmer lines. However, relying only on SINFONI data and the \emph{N2} metallicity diagnostic, without any Balmer ratios, we obtain a slightly higher global metallicity (\met = 8.94$\pm$0.07) but that it is still compatible with what is derived using only \emph{Ne3O2} and \emph{O2}. The proximity of \Ha\, and \nii\, makes the differential dust attenuation between these two lines small enough that it is still reliable to derive metallicities not including dust correction.

\section{2D Maps in Source Plane}
\label{app:2D_source_plane}

\begin{figure*}
    \includegraphics[width=\textwidth]{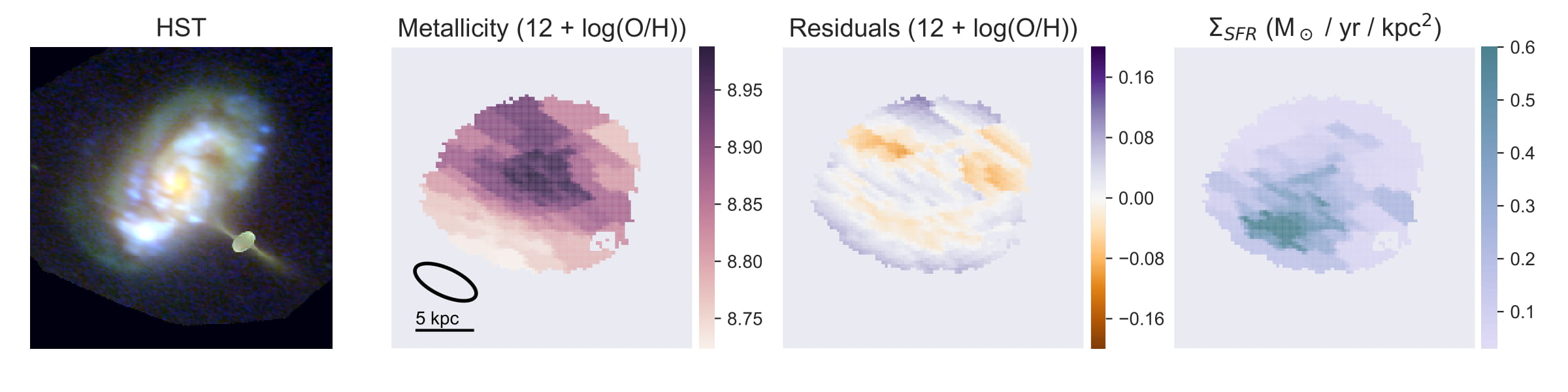}
    \caption{AS1063 in the source plane. Left: HST composite image with filters F160W, F814W and F435W. Middle Left: reconstructed metallicity map. The FWHM of the PSF in the source plane is plotted in the lower-left corner. Middle-right: source plane metallicity residuals, after subtracting the model fitted with all parameters free to vary. Right: SFR surface density map. SFRs were derived from \Hb and the \citet{Kennicutt1998} calibration.}
    \label{fig:as1063_source_plane}
\end{figure*}

\begin{figure*}
    \includegraphics[width=\textwidth]{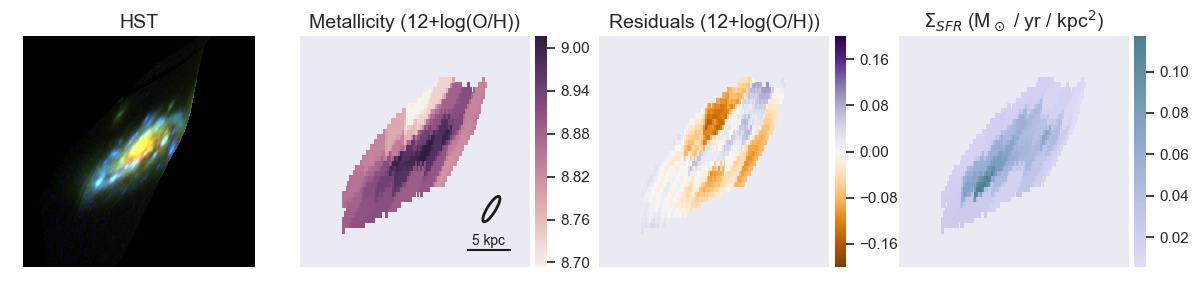}
    \includegraphics[width=\textwidth]{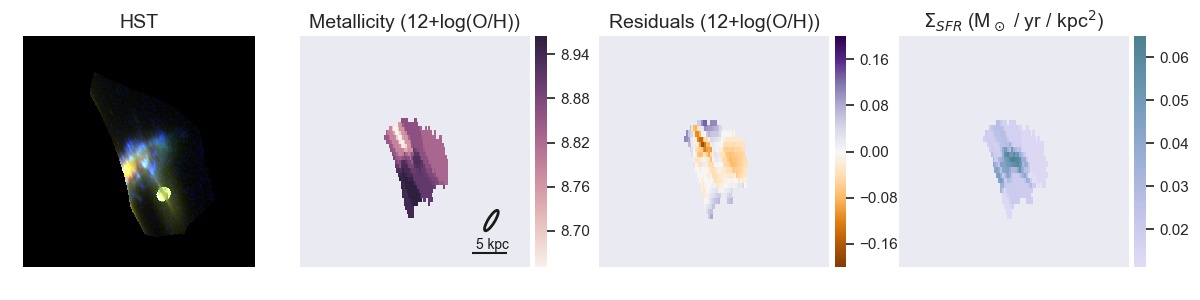}
    \includegraphics[width=\textwidth]{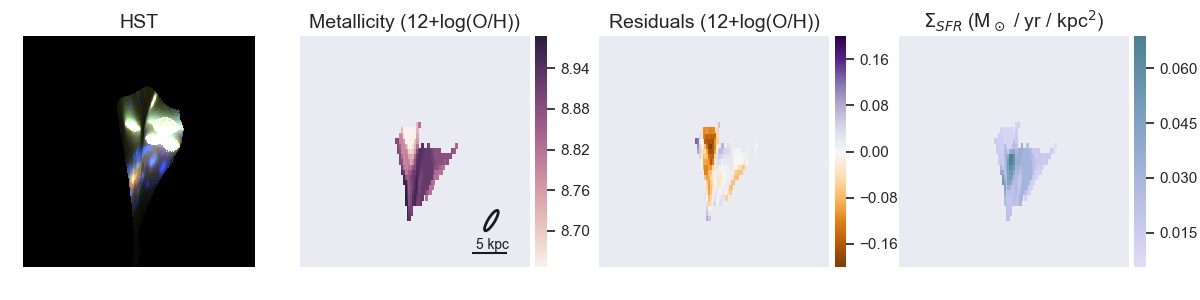}
    \caption{Same as Fig.~\ref{fig:as1063_source_plane} but for A370-sys1. Top panels are the reconstructed complete image, middle panels region 3 and bottom region 1. Region 2 is not shown due to the small area of the full galaxy it covers.}
    \label{fig:a370_source_plane}
\end{figure*}

\begin{figure}
    \includegraphics[width=0.5\textwidth]{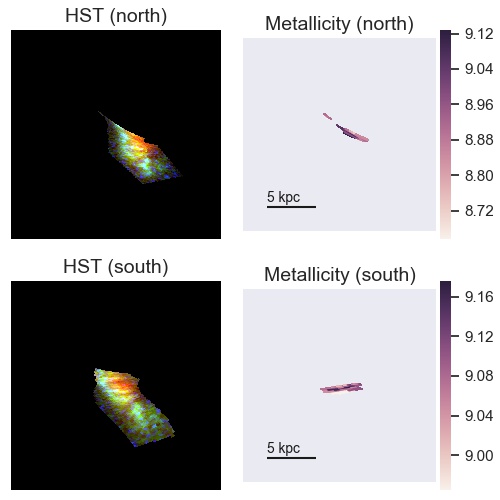}
    \caption{M1206-sys1 in the source plane. Top panels are the reconstructed multiple image in to the north and bottom panels the reconstructed image to the south.}
    \label{fig:macs1206_source_plane}
\end{figure}

We use {\sc lenstool} to correct the image plane maps of metallicity, extinction and SFR densities for lensing distortions and plot the results in Figures~\ref{fig:as1063_source_plane},~\ref{fig:a370_source_plane} and ~\ref{fig:macs1206_source_plane}. For A370-sys1 and M1206-sys1, we reconstruct the different multiple images separately. We can see that in the case of A370-sys1 (Fig.~\ref{fig:a370_source_plane}) the results from the different multiple images are sightly different, as it is expected since they come from different voxels in the data cube, but show a global agreement, with higher metallicities, E(B-V) and SFRs in the centre of the galaxy. AS1063-arc also displays higher metallicity and E(B-V) values at the centre of the galaxy. However, E(B-V) is also high in the region of higher star-formation rates, at the edge of a spiral arm.

\bsp	
\label{lastpage}
\end{document}